\documentclass[aps,prb,twocolumn]{revtex4}
\usepackage{graphics,amsmath}
\begin{document}
\title{Optical properties of small polarons from
  dynamical mean-field theory.}

\author{S. Fratini}

\affiliation{Laboratoire d'Etudes des Propri\'et\'es Electroniques des
  Solides - CNRS\\ BP 166, F-38042 Grenoble Cedex 9, France}

\author{S. Ciuchi} 

\affiliation{Istituto Nazionale di Fisica della Materia and 
Dipartimento di Fisica\\
Universit\`a dell'Aquila,
via Vetoio, I-67010 Coppito-L'Aquila, Italy}

\begin{abstract}
The optical properties of  polarons  are
studied in the framework of the Holstein model by applying
the dynamical mean-field theory. 
This approach allows to enlighten important quantitative and qualitative
deviations from the limiting treatments of small polaron theory, that should be
considered when interpreting experimental data. 
In the antiadiabatic regime, 
accounting on the same footing for
a finite  phonon frequency and a finite electron bandwidth 
allows to address the evolution of the optical 
absorption away from the well-understood 
molecular limit.
It is shown that the width of the
multiphonon peaks in the optical spectra depends 
on the temperature and on the frequency in a way that contradicts the 
commonly accepted results, most notably in the strong coupling case. 
In the adiabatic regime, on the other hand, the present method allows
to identify a wide range of parameters of experimental interest, where
the electron bandwidth is comparable or larger than the broadening of the
Franck-Condon line, leading to a strong modification of both the
position and the shape of the polaronic absorption. 
An analytical
expression is derived in the limit of vanishing  broadening, which
improves over the existing formulas and whose validity extends to any
finite-dimensional lattice. 
In the same adiabatic regime,
at intermediate values of the interaction strength, the optical absorption
exhibits a characteristic reentrant behavior,  with 
the emergence of sharp features upon increasing the temperature --- 
polaron interband transitions ---  
which are peculiar of the  polaron crossover, and 
for which analytical expressions are provided.
\end{abstract}
\date{\today}
\maketitle

\section{Introduction}

The motion of electrons in solids is often coupled to the lattice 
degrees of freedom. There are broad classes of
compounds  where the electron-lattice interaction 
is such that the carriers can form small polarons, i.e. 
they are accompanied by a local lattice
deformation that strongly modifies their physical properties, and can
lead to the self-trapping phenomenon.
This manifests in several experimentally accessible
quantities\cite{Austin}.  
First, the d.c. conductivity   is  thermally activated in a broad temperature
range, with a gap related to the energy of the electron-lattice bound
state\cite{Holstein,PRLrhopolaron}. 
This is the energy barrier that the particle has to overcome to hop from site
to site, which is proportional to the coupling strength. 
Second,  most of the   single-particle spectral weight --- a quantity 
which is experimentally
accessible through  tunneling or photoemission spectroscopy 
--- moves to high energies, corresponding to 
multi-phonon excitations in the polaron cloud
\cite{AlexRann,Mahan,sumi,depolarone,Allen}. 
Related to this, a broad contribution emerges in the optical conductivity,
typically  in the mid-infrared region, which can be
ascribed to transitions inside the polaron potential-well
\cite{old,Reik,Emin93,Mahan,Firsov,ray}.

Besides the strength, the range and specific mechanism of 
electron-phonon coupling, the phenomenon of polaron formation 
depends on the properties of the host lattice, such as the
dimensionality, the
width of the conduction band and the frequency of phonon vibrations.
In extremely narrow band systems, for example, 
the phonon energy can be of the order or even larger than the
electronic bandwidth. This leads to the concept of antiadiabatic 
quasiparticles, that can move
through the lattice being accompanied by a very fast phonon cloud,
which is at the very basis of the
standard small polaron treatments 
\cite{Holstein,Lang-Firsov}. 
In this framework,  the small polaron 
optical absorption 
is strongly reminiscent of the behavior of electrons in a gas of  independent 
molecules, and  consists 
of a series of narrow peaks at multiples of the phonon frequency, 
whose distribution is determined by the electron-phonon coupling
strength.

However, narrow band materials exist where the phonon frequency is
appreciably smaller than the electronic bandwidth, which 
therefore cannot be
described within the antiadiabatic approximation.
In this case, one has to face the effect of the relatively large 
transfer integrals between molecules, which eventually wash out 
the discrete nature of the  excitation spectra.
Contrary to the antiadiabatic situation,
where  polaron formation occurs through a smooth crossover, in the
adiabatic case a sharp transition 
separates the weak-coupling regime from the polaronic regime in dimensions
greater than 1 \cite{Kabanov-Mashtakov,Fehske-momentum,xover}.
In the weak coupling limit, the
optical absorption at low temperatures 
consists of an asymmetric band 
resulting from the excitation and absorption of a single phonon. 
In the strong coupling regime, the absorption spectrum 
moves to higher energies, reflecting the localized nature of the polaron.
In this regime, provided that the free-electron bandwidth is small
compared to the broadening of the Franck-Condon line,
the optical  absorption acquires a typical gaussian
lineshape. \cite{old,Reik}  
In the opposite situation,
i.e. when the electron dispersion dominates over the phonon-induced
broadening,  a different absorption mechanism sets in, corresponding to the
photoionization of the polaron towards the free-electron
continuum.\cite{Firsov,ray}  To our knowledge, there is at present  no
satisfactory theory of the optical spectra in the regime where both
mechanisms coexist, which is often the case in real experimental systems.
Another important unsolved question, that cannot  be addressed by
standard methods, is how
the optical absorption evolves between the weak and the strong coupling 
limit, and recent 
calculations\cite{BRoptcond,Fehske-opt} have shown that 
the spectra in the intermediate coupling regime exhibit specific
features that are characteristic of the polaron crossover region. 

The simplest model which describes the rich phenomenology depicted
above is the Holstein model \cite{Holstein}, where tight-binding
electrons are coupled 
locally to dispersionless  bosons. 
The Holstein polaron problem can be solved\cite{depolarone,sumi} 
in a non-perturbative framework
using the dynamical mean field theory \cite{RMP} (DMFT).
While this method
neglects the spatial dependence of the 
electron self energy, it
treats \textit{exactly}
the local dynamics, which makes it particularly  well
suited to describe small polaron physics. It goes  
beyond the standard analytical approaches, as it can deal 
on the same footing with the low energy effective quasiparticles \textit{and}
the high energy incoherent features, which is crucial to the
understanding of the polaron formation process\cite{depolarone}. 
The  effects of a 
nonvanishing electron bandwidth and a nonvanishing phonon frequency
are naturally included, which provides  a unified description of the 
optical  properties of small polarons, with no restrictions on the 
regimes of parameters. Compared to fully 
numerical methods\cite{ray,Fehske-opt,Fehske}, 
which are also able to tackle the intermediate coupling regimes, the
present approach is advantageous for the calculation of the optical
conductivity, since it is free of finite size effects, and 
gives direct access to the electronic excitation spectrum in real frequency. 
In contrast, finite cluster diagonalization studies suffer from the 
discretization of the Hilbert space, which can be quite severe in the 
polaronic regime where many phonon states are needed, while 
quantum Monte Carlo treatments usually work in
imaginary time, and rely on analytical continuation algorithms
for  the extraction of spectral and optical properties. 
Note however that the present DMFT results, which are based on an 
exact continued fraction expansion in the low density limit,\cite{depolarone}  
cannot be directly generalized to finite densities.

The aim of this work is to take advantage of the DMFT
to address the regimes of parameters not
covered by the standard formulas of small polaron theory. 
Applying this non-perturbative method, we can critically examine 
the ranges of validity of the usual limiting approximations, and
point out the quantitative and qualitative
discrepancies arising in several regions of the parameters space. 
Special emphasis is
given to the following points, which are not accessible by the
usual methods available in the literature: i) the 
evolution of the multi-peaked spectra in the
antiadiabatic regime, at finite values of the free-electron bandwidth; ii)
the effects of a finite electron dispersion on  the usual
Franck-Condon lineshapes, in the adiabatic polaronic  regime; iii) the
peculiar features arising in the  region of the
adiabatic polaron crossover. Concerning the last point, 
the results of our previous
work Ref.\cite{BRoptcond}, where such features were first reported, 
are here extended to much lower temperatures, which
is made possible by a new adaptive method introduced  
to deal with the fragmented excitation spectra characteristic of the 
polaron problem.  \cite{depolarone} 
This new procedure allows to identify a reentrant behavior 
of the optical properties in the crossover regime, 
where increasing the temperature switches from a weak-coupling 
like absorption to a typical polaronic lineshape.
Furthermore, a previously published formula for the
$T=0$ absorption in the polaron crossover regime is corrected here,
and analytical expressions are derived to describe
the lineshape of the polaron interband transitions  
reported in Ref.\cite{BRoptcond}.

The paper is organized as follows:
In Section II we introduce the DMFT formalism
and the Kubo formula for the optical conductivity, and discuss the
details of the calculations. Sections III and IV  are devoted to 
the analysis of the absorption spectra respectively in the
antiadiabatic  and in the adiabatic regimes. 
The  specific features arising  at intermediate values of 
the coupling strength in the  adiabatic polaron crossover region 
are examined in Section V. The main results are summarized in Section VI.

\section{Model and formalism}

We study the Holstein hamiltonian, where tight-binding electrons 
($c_{i,\sigma}, c_{i,\sigma}^\dagger$) 
with hopping amplitude $t$ are coupled locally to Einstein bosons 
($a_i,a_i^\dagger$) 
with energy $\omega_0$:
\begin{eqnarray} 
H & = &  \omega_0 \sum_{i} a^+ _{i} a_{i}  -g
\sum_{i,\sigma} c^+ _{i,\sigma} c_{i,\sigma}
 (a^+ _{i}+a_{i})  \nonumber \\
& - & t \sum_{i,j,\sigma} \left
(c^+ _{i,\sigma} c_{j,\sigma}  + {\rm H.c.} \right ). 
\label{Holstein}
\end{eqnarray}
The single polaron problem can be solved in the framework of the
DMFT\cite{RMP}, which yields an  analytical expression for
the local  self-energy $\Sigma(\nu)$ in the form of a continued fraction
expansion\cite{sumi,depolarone}. 
The latter must be iterated in an appropriate self-consistent
scheme, where the free-electron dispersion defined by the
tight-binding term in Eq. (\ref{Holstein}) enters only through the
corresponding density of states (DOS) $N(\epsilon)$. 
From the knowledge of the self-energy, one can define the single 
particle spectral function 
\begin{equation}
\label{eq:spec-func}
 \rho(\epsilon,\nu)=-\frac{1}{\pi} Im  \frac{1}{\nu-\Sigma(\nu) -\epsilon}
\end{equation}
that carries information on the spectrum of excited states,
and its momentum integral, the spectral density $N^*(\nu)$. The latter
can be evaluated by introducing the Hilbert transform of the DOS,
$ {\cal H}(z)= \int d\epsilon N(\epsilon)/(z -\epsilon)$ as:
\begin{equation}
\label{eq:spec-dens}
N^*(\nu)=-(1/\pi) \ Im \ {\cal H}[\nu-\Sigma(\nu)].
\end{equation}

The conductivity at finite frequency is related to the current-current
correlation function through the appropriate Kubo formula.
In DMFT, due to the absence of vertex
corrections\cite{DINF-OC,RMP}, such two-particle response function 
can be expressed as a functional of  the 
fully interacting single particle  spectral function 
$\rho(\epsilon,\nu)$ (or, equivalently, of the local self-energy $\Sigma$).  
This can be related to the single polaron solution of Ref. \cite{depolarone} 
by performing an expansion in the inverse fugacity \cite{BRoptcond}
$z^{-1}=\exp(\mu/T)$.
($z^{-1}\to 0$ as 
the chemical potential $\mu \to
-\infty$ in the limit of vanishing density at any given
temperature). 
The optical conductivity turns out to be proportional to the carrier
concentration $x$, and can be expressed in compact form as
\begin{equation}
\label{sigma-compact}
    \sigma(\omega)=
    \frac{x \zeta \pi}{\omega} (1-e^{-\beta\omega}) \frac{{\cal D}(\omega,\beta)}
    {{\cal N}(\beta)} 
\end{equation}
where the constant $\zeta=e^2a^2/\hbar v$ carries 
the appropriate dimensions of conductivity ($a$
being the lattice spacing, $v$ the volume of the unit cell) and
$\beta=1/T$ is the inverse temperature. 
We have defined
\begin{eqnarray} \label{caldi}
   {\cal D}(\omega,\beta) &=& 
 \int d\epsilon N(\epsilon) \phi(\epsilon) \times \\
&&\times   \int d\nu  e^{-\beta(\nu-E_0)} 
          \rho(\epsilon,\nu) \rho(\epsilon,\nu+\omega) \nonumber \\
 \label{eq:norma}
  {\cal N}(\beta)&=&\int d\nu  e^{-\beta(\nu-E_0)} N^*(\nu)
 \label{calenne}
\end{eqnarray}
where $E_0$ is the polaron ground state energy at $T=0$. The exponential
factors account for the thermal occupation of the electronic levels
(the Fermi temperature vanishes in the low density limit, and 
the particles obey Maxwell-Boltzmann statistics).  
The normalization factor ${\cal N}$ represents the partition function of the
interacting carrier. Unless differently specified, 
the density of states $N(\epsilon)$  of the
unperturbed lattice, that enters in the definition of the correlation
function ${\cal D}$, is assumed semi-elliptical of half bandwidth $D$
\begin{equation}
N(\epsilon)=\frac{2}{\pi D^2}\sqrt{D^2-\epsilon^2}
  \label{def-semicirc}
\end{equation}
corresponding to a Bethe lattice in the limit 
of infinite connectivity. This choice
reproduces the low energy behavior of a three-dimensional
lattice, while leading to tractable analytical expressions. The function
$\phi(\epsilon)$ is the corresponding current vertex 
\begin{equation}
  \label{def-vertex}
\phi(\epsilon)=(D^2-\epsilon^2)/3  
\end{equation}
 that can be derived by enforcing a 
a sum rule for the total spectral weight
\cite{Freericks,Millis}. Compared to the more rigorous
procedure of Ref. \cite{Vandongen}, this has the advantage of leading
to the correct threshold behavior for the optical conductivity 
expected in a three-dimensional system in the weak coupling
limit,\cite{Mahan} as well as in the adiabatic photoionization
limit,\cite{Emin93,Firsov,ray}  
as will be shown below. 
In the usual strong coupling regime,\cite{old,Reik} 
on the other hand, the results become 
independent on the detailed shape of both $N$ and $\phi$, 
and the choice of the vertex function is not influent. 
Let us mention that
low-dimensional systems can also be studied with the
present method, by replacing $N(\epsilon)$ and $\phi(\epsilon)$ of
Eqs. (\ref{def-semicirc}) and (\ref{def-vertex})
with the appropriate DOS and vertex function. In such cases,
the neglect of vertex corrections in the current-current correlation
function that is implicit in the DMFT formulation  could in principle 
constitute an important limitation. 
In practice, however, the limiting cases
of weak and strong coupling are correctly reproduced by the DMFT,
which gives some confidence on the results obtained in more general
situations (see e.g. the discussion at the beginning of Section IV A).

With the present 
choice of Eqs. (\ref{def-semicirc}) and (\ref{def-vertex}), 
the Hilbert transform ${\cal K} (z)$ of the product
$N(\epsilon)\phi(\epsilon)$ can be expressed in closed form, which
allows the $\epsilon$-integral in Eq. (\ref{caldi}) to be 
performed explicitely:
\begin{equation}
  \label{numerica}
  {\cal D}= \int d\nu e^{-\beta(\nu-E_0)}  
        B[\omega+\nu-\Sigma(\omega+\nu),\nu-\Sigma(\nu)]
\end{equation}
with
\[
  B(z_1,z_2)= -\frac{1}{2\pi^2} 
Re \left[ \frac{{\cal K} (z_1)-{\cal K} (z_2)}{z_2-z_1}
-\frac{{\cal K} (z_1^*)-{\cal K} (z_2)}{z_2-z_1^*}\right].
\]
The remaining integral in $\nu$ appearing in Eq. (\ref{numerica}) has to be 
computed numerically, which presents two main difficulties.
First of all, at temperatures $T\ll \omega_0$, the function $B\propto
Im \Sigma(\nu)$ 
decays exponentially at energies $\nu$ below the ground
state $E_0$. Nevertheless, such regions contribute 
to the final result due
to the presence of the thermal factor $e^{-\beta(\nu-E_0)}$. As a rule
of thumb, at low temperatures,
the integration limits must be extended down to $\nu-E_0\simeq -2E_P$
to account for a proper number of excited states  
\footnote{It is known from the atomic limit that, 
once  multiplied by   $e^{-\beta(\nu-E_0)}$, the spectral weight at negative
frequencies has a distribution
centered at $\nu-E_0\approx -E_P$
--- see Ref. \cite{Mahan}, and Fig. \ref{fig:dos-interm}}. 
At the bottom of the integration
region, the number $B$ is therefore multiplied by an exponentially 
large factor $e^{2 E_P/T}$ which amplifies the numerical errors and,
for typical 
values of $E_P/\omega_0\sim 10$, limits the attainable low temperature
limit to $T/\omega_0\sim 0.1$. The second difficulty is
that, in the strong coupling regime, the function $B$ is
zero almost everywhere for $\nu<E_0$, being concentrated in 
exponentially narrow peaks of width $\sim e^{-E_P/\omega_0}$, corresponding to
multi-phonon resonances at multiples of $\omega_0$ (see  Fig. 
\ref{fig:dos-interm}
below). For example, taking the parameters of
Fig. \ref{fig:adiab}.b, the ratio between the  width of the peaks and their 
separation is $10^{-3}$. Special care must be taken 
to manage with this rapidly varying function. 
A  uniform mesh discretization of the integral
appearing in Eq. (\ref{numerica}) was used in
Ref. \cite{BRoptcond}, which was appropriate for
the intermediate coupling/temperature regimes.
To attain the strong coupling and low temperature regimes,
we use here an adaptive non-uniform mesh optimized to 
account for the narrow peaks of the function $e^{-\beta (\nu-E_0)} Im
\Sigma(\nu)$. 

Bearing these limitations in mind, 
we shall present  in the following Sections 
the numerical results for coupling strengths up to
$E_P/\omega_0\sim 10$ and temperatures down to $T/\omega_0\sim
0.01-0.1$ (the lowest temperatures are reached in the weak/intermediate 
coupling regime). 
Analytical formulas will be derived to access the limits  
of strong coupling  and vanishing temperatures. 
For simplicity,  we shall
drop the numerical prefactor $x\zeta$ in Eq. (\ref{sigma-compact}) 
and  express all energies in units of the half-bandwidth $D$.

\section{Optical conductivity in the antiadiabatic regime}
\begin{figure*}[htbp]
\centerline{\resizebox{8.2cm}{!}{\includegraphics{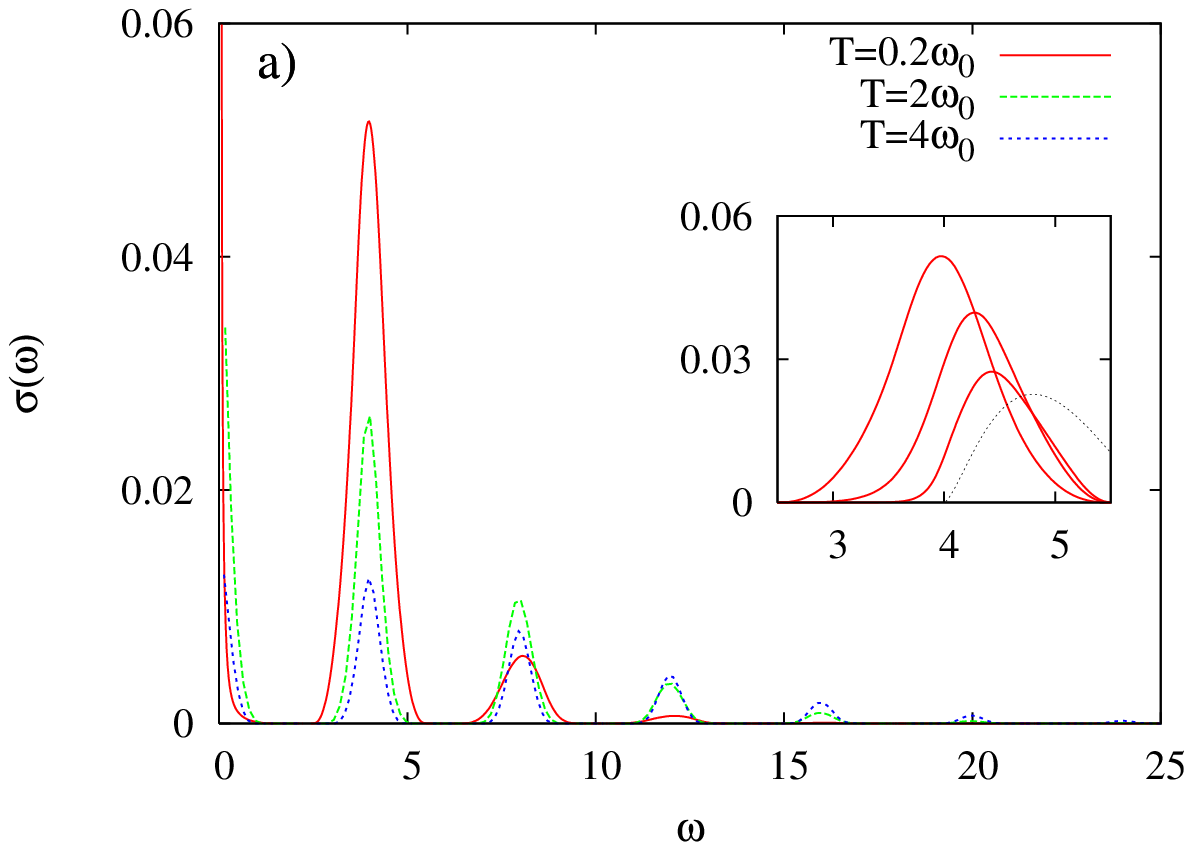}}
\resizebox{8.2cm}{!}{\includegraphics{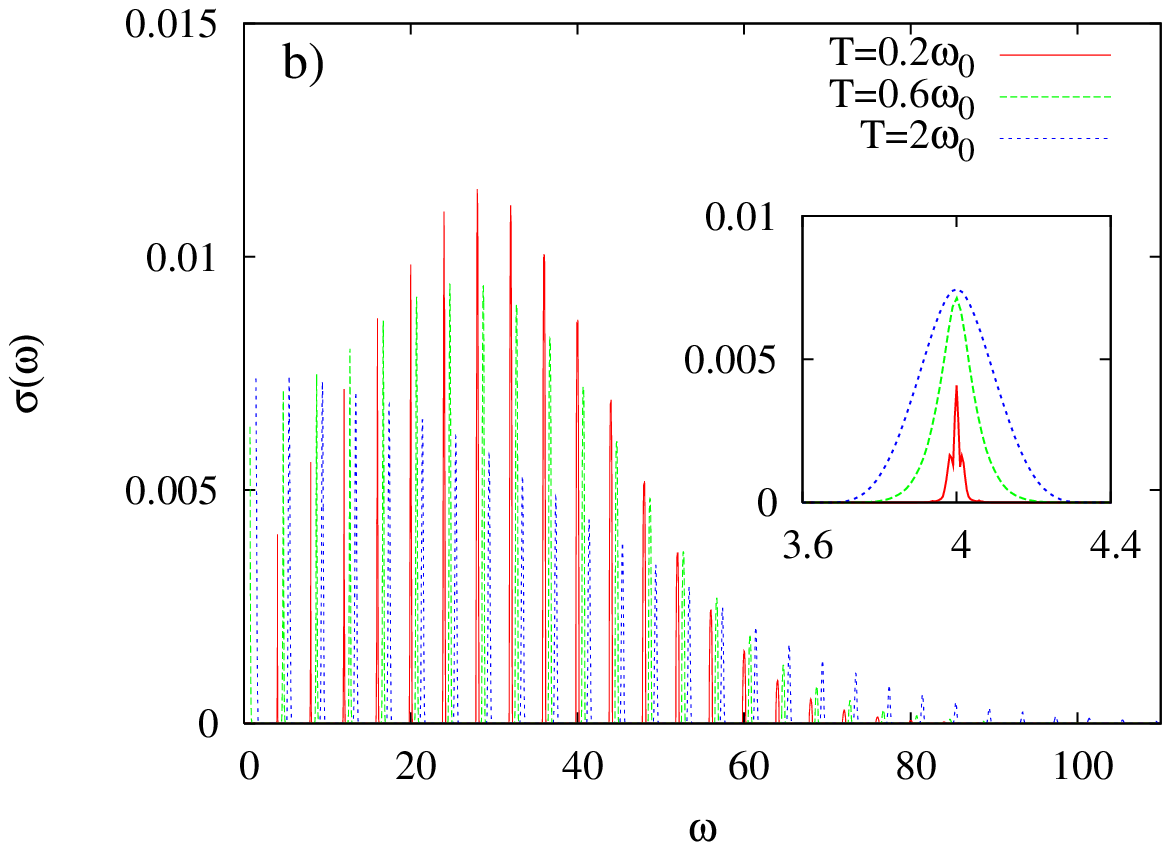}}}
\caption{(Color online) 
Optical conductivity in the antiadiabatic regime, a) for
  $\gamma=\omega_0/D=4$ and  $\alpha^2=0.25$
  and b) $\alpha^2=5$. a) In the weak coupling regime, the individual 
  absorption peaks shrink upon increasing the temperature above $T\sim
  \omega_0/2$, as predicted by
  Holstein's approximation. In the inset, the low-temperature
  evolution of the one-phonon absorption is shown (full red lines,
  from left to right,  $T/\omega_0=0.2, 0.05, 0.02$). The black dotted line
  is the weak coupling result (\ref{eq:weak-T0}) at
  $T=0$. The absorption edge shifts from  $\omega=\omega_0-2D$
  at $T\gtrsim D$ to $\omega=\omega_0$ at $T\ll D$ (see appendix A).
     b)  In the strong
  coupling regime, the optical absorption consists of extremely
  narrow peaks, similar to the case of an isolated molecule  
  (the peaks at $T=0.6\omega_0$
  and $T=2\omega_0$ have been shifted
  laterally by a constant offset for clarity).  Contrary to
  the weak coupling limit, here the individual peaks broaden with 
  temperature  (see inset).
\label{fig:anti}}
\end{figure*}

\label{sec:anti}

The process of polaron formation at zero temperature is different
depending on the value of the 
adiabaticity ratio $\gamma=\omega_0/D$,  
which measures the relative kinetic energies of phonons and band
electrons ($D$ is half the free-electron bandwidth, proportional to
the hopping amplitude $t$). 
In the antiadiabatic regime ($\gamma\gg 1$), 
the buildup of electron-lattice correlations occurs
through a smooth crossover controlled by the coupling parameter 
$\alpha^2=E_P/\omega_0$,  \cite{depolarone,xover}  
which sets the average number of phonons in the polaron cloud ($E_P\equiv
g^2/\omega_0$ is defined as the polaron binding energy on an isolated 
molecule). In this case
the behavior of the optical absorption at any value of the
electron-phonon coupling strength can be deduced 
to a good approximation from the analysis of an isolated molecule,
corresponding to the limit $D\to 0$.
\cite{old,Reik,Mahan} 
It consists of a
series of narrow peaks at multiples of  $\omega_0$, whose
distribution is determined by the coupling parameter $\alpha^2$,
and evolves  gradually through the polaron crossover located around 
$\alpha^2\sim 1$  (see also Fig. \ref{fig:PD} in the
following).  At low $\alpha^2$, the
spectral weight is mainly located in the Drude peak, plus
a weaker absorption peak at $\omega=\omega_0$. 
Upon increasing the coupling strength,
multiphonon scattering processes become important, and several
peaks  arise at multiples of $\omega_0$. 
For $\alpha^2\gg 1$, the distribution of the peak weights eventually tends 
to a gaussian centered at $\omega= 2E_P =2\alpha^2 \omega_0$, with a variance
$\sim \sqrt{2 E_P \omega_0}$ [cf. App. A and Eq. (12) below].

Thermal effects lead to some redistribution of spectral weight among
the different peaks:
In the weak coupling regime, 
the thermal  excitation of phonons
generates additional absorption peaks at multiples of  
$\omega_0$.
In the strong coupling regime, increasing
the temperature  broadens the 
distribution of peak weights: For $T\gtrsim \omega_0/2$, the
variance  tends to $\sqrt{4 E_P T}$ and is  completely 
determined by the thermal fluctuations of
the phonons  (cf. App. A); 
When the temperature is
increased further, thermal dissociation of the polaron at $T\sim E_P$
eventually washes out the absorption maximum at finite frequency, 
resulting in a transfer of spectral weight towards $\omega=0$.    

The above  picture is based on the limit of
vanishing electron bandwidth. A small nonvanishing transfer integral
can be expected to give the individual peaks a finite width, 
without modifying the distribution of spectral weights \cite{AlexRann}.
In the weak coupling regime, 
this scenario is confirmed by the DMFT results
illustrated in Figure
\ref{fig:anti}.a  for $\gamma=4$ and $\alpha^2=0.25$. The absorption spectrum
at $T=0.2 \omega_0$ consists of few peaks of width $\simeq 4D$, centered 
at multiples of $\omega_0$; Increasing the temperature 
leads to a moderate transfer of spectral weight to higher
frequencies, while the individual peaks 
shrink  as predicted by Holstein's approximation\cite{Holstein}. An interesting
behavior is seen at extremely low temperatures, where $T\ll
D\ll \omega_0$ (inset of fig 1.a). In this limit, the optical
absorption is dominated by transitions between ``low momentum''
states, i.e. states located near the
bottom of the subbands. As a result, the peaks become asymmetric and
the absorption thresholds move from $\omega=n\omega_0-2D$ to
$\omega=n\omega_0$ 
[details are given in Appendix A, Eq. 
(\ref{eq:weak}) and (\ref{eq:weak-anti})].

The optical conductivity in
the strong coupling regime  ($\gamma=4$ and $\alpha^2=5$) is shown in
Fig. \ref{fig:anti}.b. In this case, although the overall picture
deduced from the molecular limit is qualitatively recovered, 
the fine structure 
exhibits a behavior that  contradicts the standard results available in the
literature\cite{Holstein,Reik,AlexRann,Lang-Firsov},
according to which the absorption spectra at nonzero $D$ consist of
a series of narrow peaks of equal width, proportional to the
renormalized polaronic bandwidth [this follows, through
Eq. (\ref{caldi}), from the assumption that the multiphonon subbands 
in the single particle  spectral density all have the 
same width].
Instead, the multiphonon peaks in Fig. \ref{fig:anti}.b are broader at 
high frequency than at low frequency,
although this is hardly visible on the scale of the figure. 
More surprisingly, their width \textit{increases} with temperature, as
shown in the inset of Fig. \ref{fig:anti}.b, contrary to
what could be expected\cite{Holstein}.

This apparently anomalous behavior  
can be traced back to finite bandwidth effects
on the excitation spectrum, that 
go beyond Holstein's decoupling scheme, and
have already been reported in
Ref. \cite{depolarone} as well as in Ref.\cite{PRLrhopolaron}, 
where they have been shown to  strongly affect the absolute value of the  
d.c.  conductivity. In fact, the broadening of molecular levels 
induced by band overlap 
is not uniform over the excitation spectrum, but rather 
increases as they move away from the groud state (see e.g. Figs. 17
and 18 in Ref. \cite{depolarone}): while the width of the  lowest
peak, related to coherent tunneling between different molecular units,
scales exponentially as $\exp(-\alpha^2)$,  
the width of the higher order peaks is much larger, being determined by both
incoherent hopping between neighboring
molecules and coherent hopping in the presence of multiphonon excited states:
for such peaks, the DMFT results indicate a power law dependence in $\alpha^2$.
Moreover, as the temperature increases, the different levels are mixed
by phonon thermal fluctuations. 
As a result, 
the width of the subbands at low energy rapidly increases with
temperature to attain the typical broadening of the high energy subbands,
as seen in the inset of Fig.  \ref{fig:anti}.b (in the data at $T=0.2\omega_0$, 
a fine structure due to the superposition of at least 
two contribtions of different widths is visible).
 
Note that the  discrete multi-peaked structure
characteristic of the  antiadiabatic regime remains at all
temperatures, contrary to what is  stated in Ref. \cite{Reik}:  
when $D\ll \omega_0$, the thermal broadening of the peaks is not
sufficient to lead to a continuous absorption curve, even at $T\gg
E_P$, 
unless a sizeable phonon dispersion is introduced in the model (this
would give the individual peaks an intrinsic finite width, 
preventing the extreme band narrowing in the limit $D\to 0$).

\section{Optical conductivity in the adiabatic regime}

In the adiabatic limit ($\gamma = 0$),
the carrier properties change drastically
from weakly renormalized electrons to self-trapped polarons 
at a critical value of  the coupling strength $\lambda\equiv E_P/D$, above
which  a bound state  emerges below  the bottom of the free-electron band.
For example, in the case of the semi-elliptical density of states Eq. (7), 
adiabatic polaron formation  takes place at
$\lambda_c=0.843$.
Allowing for lattice quantum fluctuations changes this localization transition
into a very sharp crossover (cf. Fig.  \ref{fig:PD}) 
which separates weakly renormalized electrons
for $\lambda\lesssim \lambda_c$ from polarons with very large effective mass
at $\lambda\gtrsim\lambda_c$. \cite{depolarone,xover}

To describe this behavior, methods such as the DMFT that 
do not rely on a ``small'' parameter  are extremely valuable
for the following reasons. 
On one hand,  there
is at present no unified analytical approach which allows to calculate 
the optical conductivity (even approximately)  in the whole range 
of parameters, from the weak to the strong coupling limit. 
This is not surprising,
since the physics is fundamentally different in the two regimes. This
situation should be contrasted with the antiadiabatic case, where the basic
qualitative features of the optical absorption 
can be inferred form the solution of the model on a single molecule (cf. 
beginning of Section III A).
In addition, even within the adiabatic strong coupling regime,
the polaronic lineshape changes depending on the ratio between the
broadening $s$ of electronic levels induced by phonon fluctuations,
and the free electron bandwidth $D$. Theories exist for
the limits $s/D\to 0$ and  $s/D\to \infty$, but 
their validity is questionable in the intermediate range of 
experimental relevance.

In the following, we shall first report the analytical
expressions recovered  within the  present DMFT formalism,
in three limiting cases: weak coupling, strong
coupling $s/D\to \infty$ and  $s/D\to 0$. The first two  
formulas turn out to be perfectly equivalent to the results available in
the literature, showing that at least in such limits, 
the neglect of vertex corrections in the Kubo formula implicit in the DMFT
approach has negligible influence on the results.
The third formula, on the other hand, constitutes an 
improvement over the expressions of Refs. \cite{Firsov,ray}, even when 
applied to  one-dimensional lattices.
The DMFT results obtained in more
general cases will be presented next, 
pointing out the
inadeguacy of the standard descriptions in many cases of interest.

\subsection{Limiting cases}

In the weak coupling limit ($\lambda\to 0$), 
the optical absorption at $T=0$ consists of a broad
band  related to the excitation of a
single-phonon, with an edge at $\omega=\omega_0$ 
followed by a power law  decay at higher frequencies, and eventually an upper
edge at $\omega=\omega_0+2D$
[cf. Eq. (\ref{eq:weak}) in Appendix A]
\begin{equation}
  \label{eq:weak-T0}
  \sigma(\omega)=\frac{E_p \omega_0 \pi}{\omega^3}
 \phi(\omega-D-\omega_0)
N(\omega-D-\omega_0).
\end{equation}
In a three dimensional system, the absorption edge behaves as
$(\omega-\omega_0)^{3/2}$. 
As the  temperature increases, the
gap below the threshold is rapidly filled and the  absorption maximum
is washed out  above $T\sim \omega_0/2$.

In the strong coupling regime ($\lambda\to \infty$),  the
photoexcitation of the electron is much faster than the lattice
dynamics,  which  is virtually frozen during the absorption process. 
Since the lattice energy cannot be relaxed, the dominant optical transition
corresponds to the difference in \textit{electronic} energy between
the initial and final states (Franck-Condon principle) which, 
in the  Holstein model,  equals twice the ground state energy $2 E_P$.
The shape of the optical absorption will depend on the ratio
between the width  of the non-interacting band $\sim D$,
 and the variance $s$ of the
phonon field, which controls the broadening
of electronic levels. The latter obeys  \cite{HoHu}
\begin{equation}
  \label{eq:variance}
  s^2=E_P \omega_0 \coth{\omega_0/2T}
\end{equation}
(cf. Appendix A).
It is determined by the quantum fluctuations of the
phonons at low temperatures, and increases 
due to the thermal fluctuations as $T\gtrsim \omega_0/2$.

When $s\gg D$, i.e. when the phonon induced broadening of electronic levels
is much larger than the electronic dispersion,
the absorption by localized polarons 
takes the form  of a skewed gaussian peak centered at $\omega_{max}=2E_P$:
\cite{old,Reik}
\begin{equation}
\label{eq:atomic-adiabatic-text}
  \sigma(\omega)=\frac{\pi}{\omega}\frac{D^2}{4}\frac{1-e^{-\omega/T}}
{\sqrt{4 \pi s^2}} 
\exp\left[-\frac{(\omega - 2E_P)^2}{4 s^2}\right].
\end{equation}
Following  Eq. (\ref{eq:variance}), the Franck-Condon line further 
broadens upon increasing the temperature above $T\sim \omega_0/2$ and 
eventually moves  towards $\omega=0$  at temperatures
higher than the polaron binding energy, as the polaron thermally dissociates. 
Note that the above formula also describes the \textit{envelope} of
the discrete absorption spectrum of polarons in the antiadiabatic
regime, shown in the preceding Section (cf. Fig.  \ref{fig:anti}.b). 

To recover the
standard  result for a three-dimensional cubic lattice,\cite{Mahan} where the
total bandwidth is $2D=12t$, the above formula has to be multiplied by 
a prefactor $2/9$, which corrects for the larger value of the
average square velocity in the DOS Eq. (7) used in the
calculations. Analogous prefactors should be included when any of the DMFT
results obtained in this work are applied to 
finite-dimensional lattices, 
that can be straightforwardly obtained by evaluating the second moment of the 
corresponding non-interacting DOS.

In the opposite limit $s \ll D$,  the lineshape  is dominated by the  
electronic dispersion. 
The absorption is due to transitions from a polaronic state whose
electronic energy is $\simeq -2E_P$ to the continuum of free-electron states.
We have [cf. Eq. (\ref{eq:Fratinov}) in Appendix A]
\begin{equation}
  \label{eq:fratinov-text}
  \sigma(\omega)=\pi \frac{ 4E_P^2}{\omega^2}\frac{1-e^{-\omega/T}}{\omega}
\phi(\omega-2E_P)N(\omega-2E_P)
\end{equation}
We see that in this case the  absorption of photons is only possible
in the interval $2E_P-D<\omega<2E_P+D$. In three dimensions, 
the absorption vanishes as $\Delta \omega^{3/2}$ at the edges, as in the 
weak coupling case of Eq. (10). 
Taking a semi-elliptical DOS as representative of a
three-dimensional lattice, we also find that 
the absorption  maximum $\omega_{max}=2E_P-D^2/2E_P$ 
is shifted to lower frequencies
compared to the usual estimate. This softening  is entirely due to finite
bandwidth effects. Note that Eq.(\ref{eq:fratinov-text}) is valid
at all temperatures below the polaron dissociation  temperature $T\sim
E_P$. In particular, contrary to Eq. (12),  nothing  happens here  at
temperatures $T\sim \omega_0/2$, provided that the condition  $s \ll D$
is not violated.

Formulas similar to Eq. (\ref{eq:fratinov-text}) 
have been derived in Ref. \cite{Firsov}, in Ref. \cite{ray} for
one-dimensional systems, and in Ref. \cite{Emin93} for the 
high frequency absorption threshold
of large polarons in the  Fr\"ohlich model.
As discussed in Appendix A, these derivations are strictly valid only 
in the limit $E_P\to \infty$. On the other hand, 
the present formula   accounts for 
the non-negligible dispersion  of the initial state at finite values
of $E_P$, which is reflected in the additional
prefactor $4E_P^2/\omega^2$, leading in general 
to a much more asymmetric lineshape. 
Remarkably, if the appropriate one-dimensional DOS is
used, Eq. (\ref{eq:fratinov-text}) describes much better the
one-dimensional exact
diagonalization data of  Ref. \cite{ray} than their own Eq. (9)
[also reported in Appendix A as Eq. (\ref{eq:Ray})]. 
The absorption maximum in this case is located at $\omega_{max}=
2E_P-E_P(\sqrt{1+6(D/E_P)^2}-1)/2$.

\begin{figure}[htbp]
\resizebox{7.5cm}{!}{\includegraphics{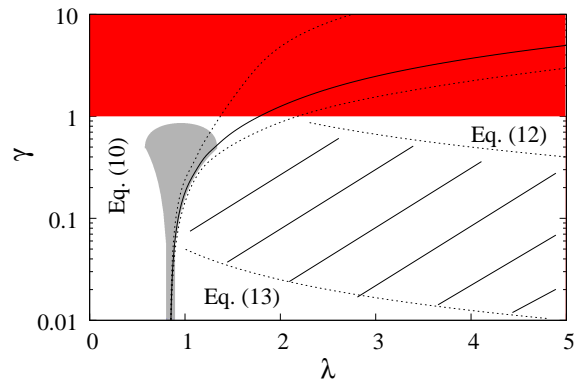}}
\caption{(Color online) Phase diagram  
  illustrating the regimes of validity, at $T=0$, of the
  different limiting formulas derived in the text. 
  The bold and dotted lines indicate the location and spread   
  of the polaron crossover as defined in Ref.\cite{xover}.
  The (red) shaded region delimits the antiadiabatic
  regime $\gamma >1$ treated in Section III. 
  In the hatched region at $\gamma <1$,  the optical conductivity 
  is not accessible by the usual methods, as pointed out  in Section
  IV.   The (gray) shaded 
  region indicates the adiabatic intermediate regime 
   described in Section V.   }
  \label{fig:PD}
\end{figure}

The regions of validity of the analytical  formulas presented in this Section 
are summarized in Fig. \ref{fig:PD}.

\subsection{DMFT results}

\begin{figure*}[htbp]
\centerline{\resizebox{8.2cm}{!}{\includegraphics{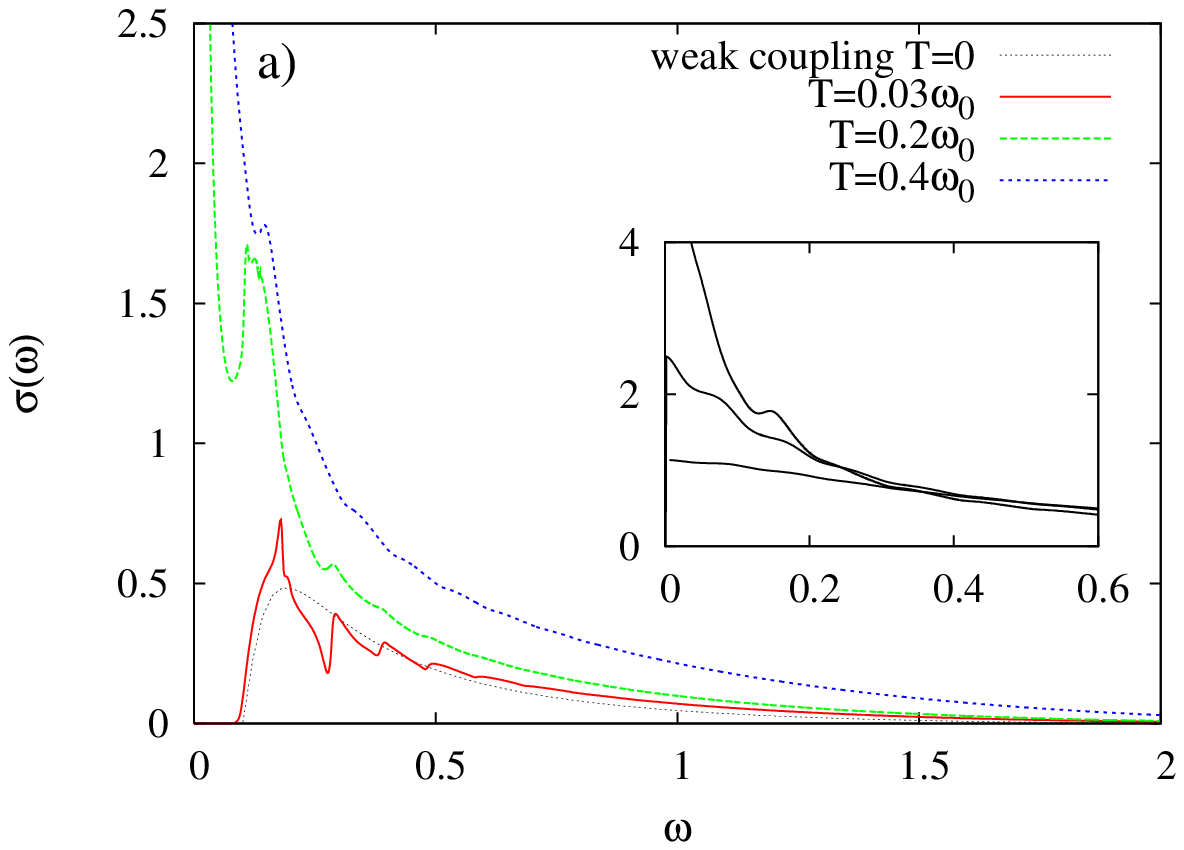}}
\resizebox{8.2cm}{!}{\includegraphics{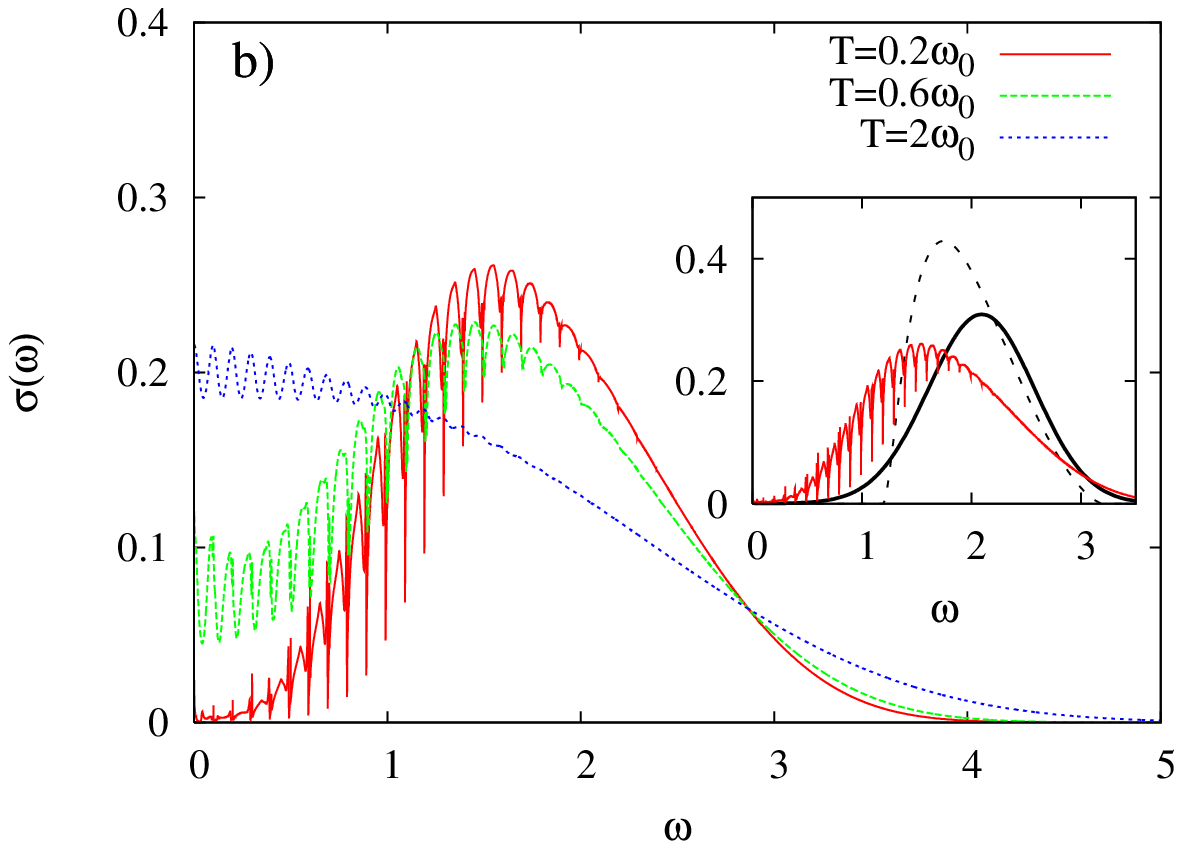}}}
\caption{(Color online) 
Optical conductivity in the adiabatic regime, a) $\gamma=\omega_0/D=0.1$
  and $\lambda=0.7$ and b) $\lambda=1.1$. a) In the weak coupling
  regime, the gap below $\omega=\omega_0$ is rapidly filled upon
  increasing the temperature, and the one-phonon absorption threshold is washed
  out at  $T\gtrsim \omega_0/2$. The black dotted line is the analytical
  weak-coupling result  (\ref{eq:weak-T0}).
  At higher temperatures, the 
  low-frequency absorption is reduced as some spectral weight is
  transferred to
  higher frequencies (see inset: from top to bottom, $T/\omega_0=0.4, 0.6$ 
  and $1$).  b) In the polaronic
  regime, the absorption peak  broadens at  $T\gtrsim \omega_0/2$ and then 
  moves towards
  $\omega=0$ as $T\gtrsim E_P$.    The  inset compares the low
  temperature spectrum with the limiting lineshapes Eq. (12) (full
  line)  and Eq. (13) (dashed line).
\label{fig:adiab}}
\end{figure*}

The DMFT results obtained in the adiabatic regime 
are illustrated in Fig.  \ref{fig:adiab}. 
First of all, our data show that 
the absorption shape  (\ref{eq:weak-T0}) derived in the
limit of $\lambda\to 0$ remains qualitatively valid 
at finite values of $\lambda$, up to
the polaron crossover.  
This is exemplified in the data reported in 
Fig. \ref{fig:adiab}.a. at $\gamma=0.1$ and $\lambda=0.7$, i.e. at 
a value of the coupling strength  which
lies {\it a  priori} well beyond the range of validity of perturbation
theory,  \cite{xover}  not far from the 
polaron crossover  coupling strength $\lambda_c(\gamma=0.1)\simeq 0.93$
(see Fig. \ref{fig:PD} and Section V below). 
Despite such relatively large $\lambda$, 
the 
threshold at $\omega_0$ and the overall behavior of the high frequency
tail at low temperatures (full red line) are quite close to the
predictions of the weak-coupling  formula (\ref{eq:weak-T0})
(black, thin dotted line). The increasing importance of multi-phonon
processes shows up mainly in the  fine structure, with the
appearance of alternating peaks and dips  at multiples of $\omega_0$.

Figure \ref{fig:adiab}.b shows the DMFT results 
at $\gamma=0.1$ and $\lambda = 1.1$. This value of the coupling
strength lies in the polaronic
regime $\lambda>\lambda_c$, in a region where the electronic
dispersion and the phonon-induced broadening are comparable
($s/D=0.33$), so that  none of Eqs. (12) and (13) is expected to  hold.
The comparison with such limiting formulas is
illustrated in the inset of Fig.  \ref{fig:adiab}.b. While the position of the absorption
maximum seems to agree with the prediction 
of Eq. (13) (dashed line), the peak height  
is much reduced and the absorption edge
is completely washed out by  phonon fluctuations. On the other hand,
the temperature dependence qualitatively agrees with Eq. (12) (full
line), i.e. the peak broadens  as the
temperature is raised above $T\sim \omega_0/2$,
which can be understood because
the low frequency tails are dominated by the phonon fluctuations.
We see that the DMFT spectrum  not only
lies at lower frequencies compared to Eq. (12), but it is also much
broader and asymmetric (the same trend was observed 
 in Ref.\cite{Fehske-opt}
in the one-dimensional case). 
All these effects can be ultimately ascribed to the
finite electron bandwidth $D$. 

Our results show that detectable deviations from
Eq. (12) arise as soon as the non-interacting  bandwidth 
is larger than the broadening $s$, 
a condition that is commonly realized in real systems. 
For example, taking typical values  
$\omega_0\simeq 0.01-0.05 eV$ and $E_p\simeq
0.1-0.5 eV$ yields a zero temperature broadening $s\simeq 0.03-0.16
eV$, in which case electron bandwidths of few 
tenths of $eV$ are already sufficient to invalidate the standard  
gaussian lineshape Eq. (12). 
In the opposite limit, 
 marked deviations  from Eq. (13) appear  already at small values of the
ratio $s/D$, roughly $s/D\gtrsim 0.05$: as soon as a small finite broadening 
is considered,   exponential tails of width $s$ arise that  
wash out  the sharp absorption edge of Eq. (13), while the peak height is 
rapidly reduced (this can be partly ascribed to a rigid 
increase of the renormalized band dispersion,  
as evidenced in 
Ref. \cite{HoHu}).
The intermediate region where both Eq. (12) and (13) fail in describing the 
polaronic optical absorption is shown as a hatched area  in Fig. \ref{fig:PD}.

As was mentioned above, the ability to describe the continuous
evolution of the absorption shape between the limiting cases of
Eq. (12) and Eq. (13) ultimately follows from the fact that  
finite bandwidth effects are properly included in the DMFT.
On the other hand,
accounting for a non-zero phonon frequency
allows to address the multi-phonon fine structure of the optical
spectra, which is seen to evolve
gradually upon varying  the adiabaticity ratio $\gamma$.
In fact, the optical absorption of a polaron at finite values of $\gamma$ 
shares features with both the antiadiabatic and adiabatic limits, 
having a pronounced  structure at low frequencies (possibly with multiple 
separate  narrow bands), followed by structureless tails at higher
frequencies (cf. Fig.  \ref{fig:adiab}.b). 
As  $\gamma$ increases, the fine structure progressively extends
to higher frequencies,  and evolves into the discrete absorption
pattern  of  Fig.  \ref{fig:anti}.b, typical of the antiadiabatic limit.

\begin{figure*}[htbp]
\resizebox{8.2cm}{!}{\includegraphics{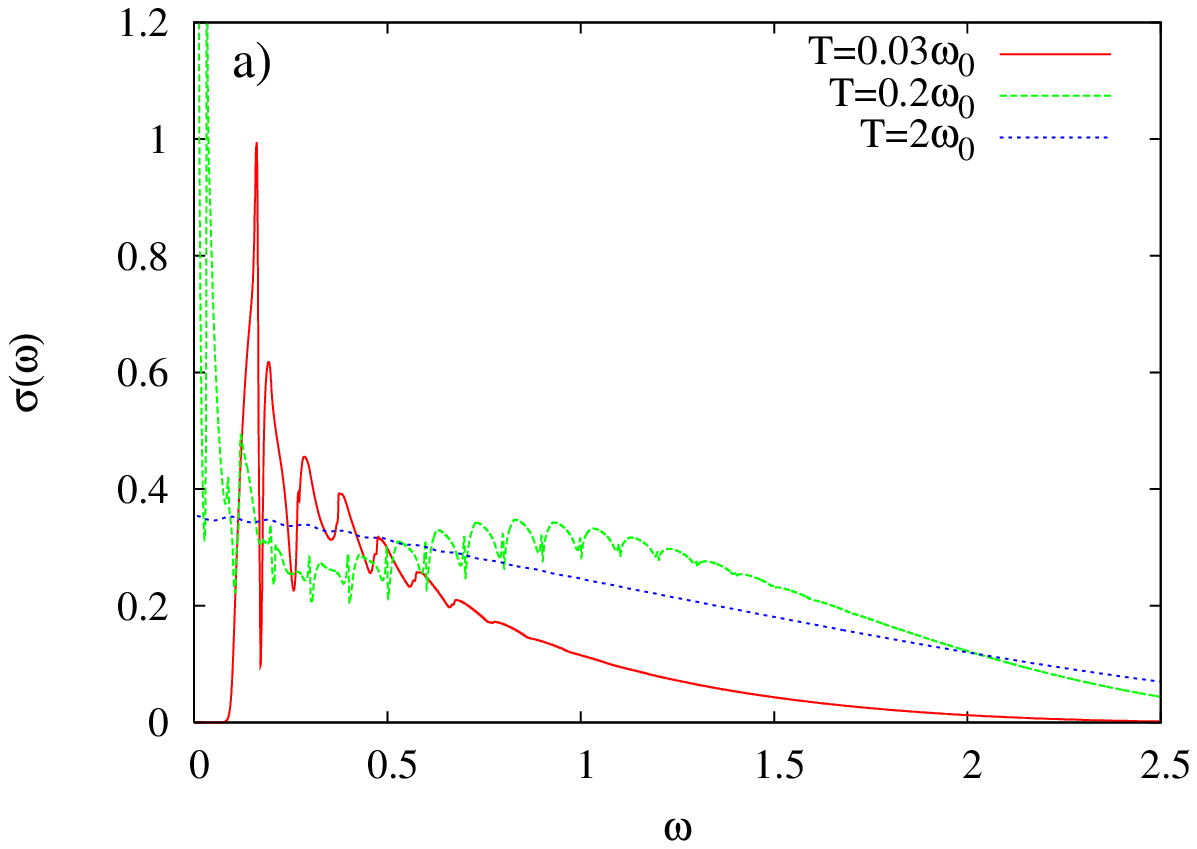}}
\resizebox{8.2cm}{!}{\includegraphics{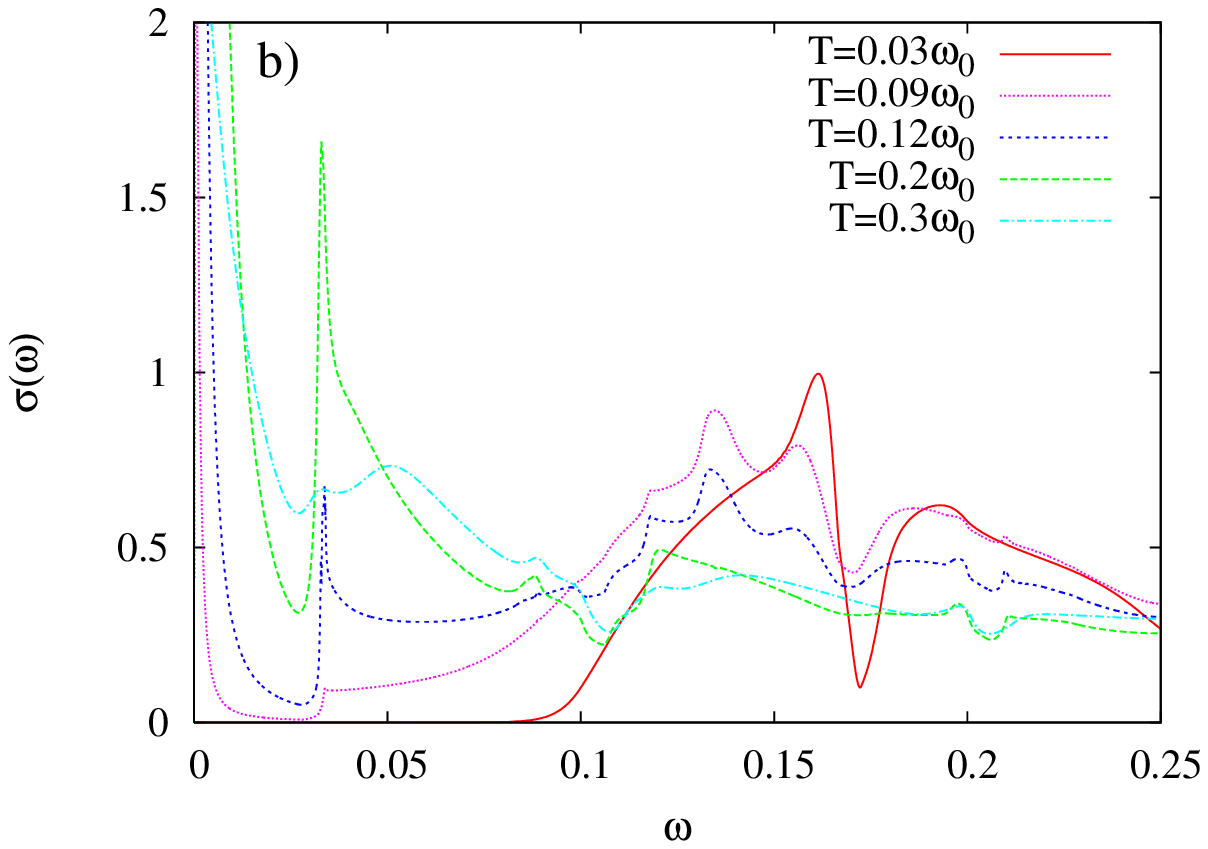}}
\caption{(Color online) 
Optical conductivity in the adiabatic regime $\gamma=0.1$, at
an intermediate value of the coupling strength $\lambda=0.9$.   
a) the absorption is ''weak-coupling like'' at low
temperatures, with a clear threshold at $\omega=\omega_0$, and becomes
''strong-coupling like'' at higher temperatures,
causing a non-monotonic evolution of the spectral weight.
b) a detailed view of the temperature evolution around the
one-phonon threshold, where Polaron Interband Transitions appear.
\label{fig:sigma-interm}}
\end{figure*}

\section{Polaron crossover in the adiabatic regime}

The critical coupling strength for polaron formation 
was identified in Ref. \cite{xover} 
as the value at which $d E_0/d g$ has maximum slope, 
which follows approximately  $\lambda_c(\gamma)\simeq 0.84 +0.9
\gamma$ at small $\gamma$  (note
that this derivative is related, through the Hellmann-Feynman theorem,
to the electron-lattice correlation function).
The width of the crossover region, obtained by looking at
the maximum slope of $|\partial^2 E_0/\partial g^2|$, is roughly given by  
$\Delta \lambda \simeq \gamma/2$ (cf. Fig. 3.b in ref \cite{xover},
reported in Fig. \ref{fig:PD} here), 
and vanishes in the adiabatic limit
where the polaron formation becomes a 
true localization transition.

In this Section, we address the specific properties of the optical absorption 
in the adiabatic polaron crossover region.
The present results extend  the preliminary   results obtained in
Ref.\cite{BRoptcond} allowing an inspection of the very low temperature regime,
which turns out to be crucial to understand the evolution of the 
optical properties.
We show that the adiabatic crossover 
region has two original signatures: The first is
a coexistence of  both weak and strong coupling characters   
in the optical conductivity, with a transfer of spectral weight
between the two occurring at very low temperatures.
The second is the emergence of narrow
absorption features at low frequencies (of the order of, or even below
the phonon frequency) with a non monotonic temperature dependence,
corresponding to resonant transitions between long-lived states in the
polaron excitation spectrum. 
\footnote{Additional peaks have also been reported to arise 
in the adiabatic intermediate coupling regime, in one space  dimension
\cite{Fehske-opt}. 
Such peaks 
have been ascribed to transitions between polaron states with different
spatial structures, and cannot be described by the present (local) 
DMFT treatment.}
We finally provide a semi-analytical expression 
for the optical conductivity in the zero temperature limit,
that corrects Eq. (7) of Ref. \cite{BRoptcond}.

\subsection{Low temperature transfer of spectral weight}

A typical  optical conductivity spectrum 
in the crossover region is shown in 
Fig. \ref{fig:sigma-interm}.a, for $\gamma=0.1$ and $\lambda=0.9$.
The data at the lowest temperature (red full curve) 
are very similar to the $\lambda=0.7$ 
results of Fig. \ref{fig:adiab}.a, with a marked
single-phonon edge at $\omega=\omega_0$ as given by
Eq. (\ref{eq:weak-T0}), followed by a multi-phonon structured 
high frequency tail. Such ``weak-coupling'' behavior is in agreement
with the fact that $\lambda=0.9$ lies slightly below the 
critical value $\lambda_c(\gamma=0.1)\simeq 0.93$. 
Upon increasing the temperature (green dashed curve), 
the optical absorption develops a 
broad maximum at high frequency typical of the polaronic
regime, comparable to what is seen at $\lambda=1.1$ 
in Fig. \ref{fig:adiab}.b.  Remarkably, this occurs at
temperatures much below the phonon quantum 
to thermal crossover at $T\sim \omega_0/2$, suggesting
that such transfer of spectral weight is governed by an electronic energy
scale. 

\begin{figure}[htbp]
\centerline{\resizebox{9cm}{!}{\includegraphics{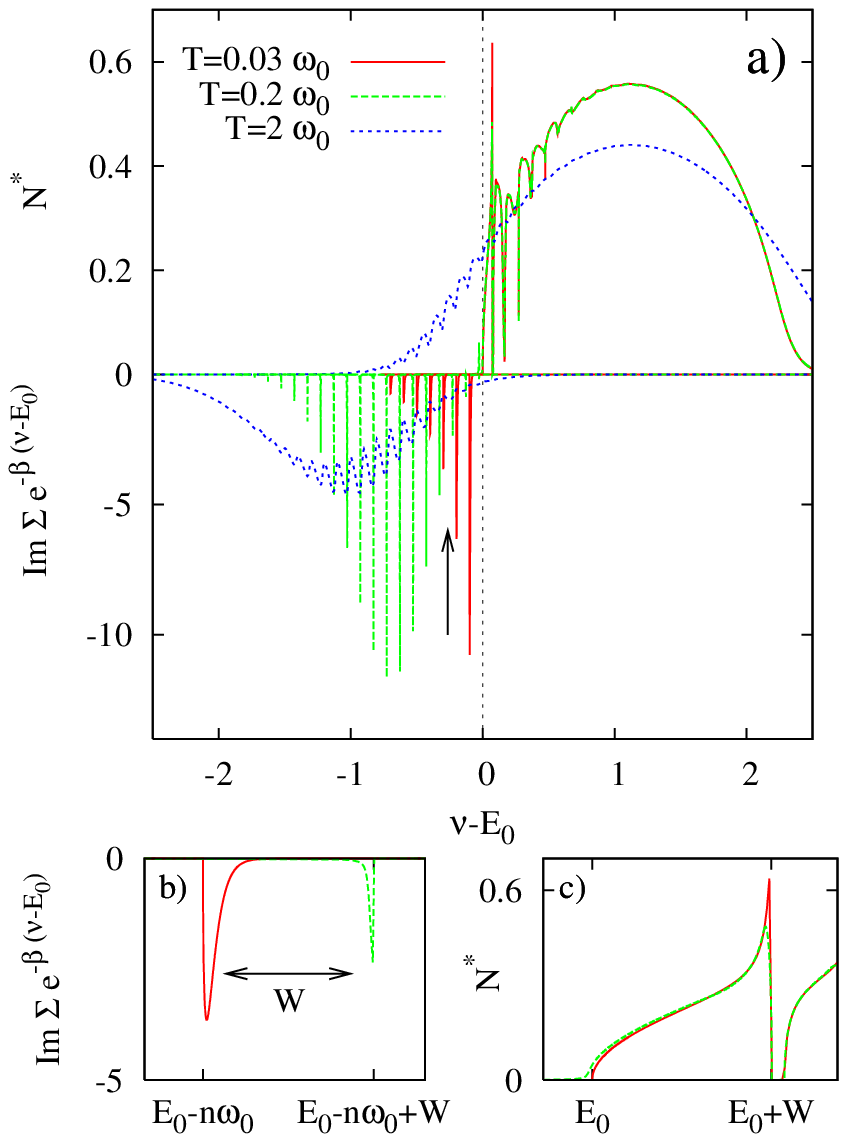}}}
\caption{(Color online) 
a) Spectral density  $N^*$ and weighted scattering rate $Im
  \Sigma(\nu) e^{-\beta (\nu-E_0)}$ at $\lambda=0.9$ and
  $\gamma=0.1$, at different temperatures. 
b) Expanded view  of a generic $n$-th peak in the
  weighted scattering rate ($n=3$, indicated by an arrow in
  the main panel, whose width equals the renormalized polaron bandwidth
  $W$  (the data at $T=0.03\omega_0$ have been multiplied by a factor $10^3$). 
The weight moves from the bottom to the top of the band at
  $T\sim W$. c) Expanded view of the polaron band in the spectral density
  $N^*$, separated from the high-energy continuum by a gap.
\label{fig:dos-interm}}
\end{figure}

To understand the origin of this phenomenon, 
a more detailed analysis of the single-particle spectral function is needed.
As can be seen in  Eq. (\ref{caldi}), the calculation of the optical
conductivity involves a convolution of  
the two quantities $\rho(\epsilon,\nu+\omega)$ and  $e^{-\beta(\nu-E_0)}\rho(\epsilon,\nu)
$. Due to the exponential weighting factor, the
contributions to $\rho(\epsilon,\nu)$ at $\nu>E_0$  are strongly suppressed at low
temperatures, and only the states related to
phonon emission processes at $\nu<E_0$ will contribute.
For such incoherent states, the spectral
function is directly proportional to the scattering
rate $Im \Sigma(\nu)$ [cf. Eq. (\ref{rom}) in
App. B]. Therefore, neglecting  the 
$\epsilon$-dependence in Eq. (\ref{caldi}), 
the optical spectrum is roughly proportional to  a convolution  of the
spectral density $N^*(\nu+\omega)$ with the
weighted scattering rate $e^{-\beta(\nu-E_0)} Im \Sigma(\nu)$.
Both of these quantities are illustrated in Fig.  \ref{fig:dos-interm}.

With the present choice of parameters, 
the spectral density at low temperature (upper panel of Fig. 
\ref{fig:dos-interm}.a) 
consists of a single subband
of width $W<\omega_0$ (the polaron band, expanded in Fig.
\ref{fig:dos-interm}.c) disconnected from the electron
continuum at higher energy. 
Scattering processes are suppressed  below
$E_0+\omega_0$ because we are considering dispersionless optical
phonons, so that 
the whole polaron band  represents 
\textit{coherent} long-lived states at low temperature. 
In contrast, higher energy states at $\nu>E_0+\omega_0$ are
mostly incoherent, being  strongly scattered by phonons.
The marked asymmetry of the polaron band in
Fig. \ref{fig:dos-interm}.c 
comes from the fact that
the band dispersion flattens in proximity of the band top, where the states
have a more localized character.
This coexistence of free-electron like states at low momenta and
localized states at high momenta
\cite{Fehske-momentum} is crucial here, 
and can be understood from the following argument:
In the weak coupling limit, 
the spectral density develops a dip at $\nu = E_0+\omega_0$ corresponding to the
threshold for one-phonon scattering. Below this value, the band dispersion 
flattens
due to the hybridization with the (dispersionless) phonon
states, leading to a very large density of states.
When $\lambda$ increases, the dip 
eventually becomes a true
gap and a polaron band of width  $W<\omega_0$ emerges from the
continuum,
retaining  a  characteristic asymmetric shape with a 
maximum close to the top edge.

The excitation spectrum  hardly changes when going from $T=0.03
\omega_0$ to $T=0.2 \omega_0$, except for the appearance of
few exponentially weak replicas of  the polaron band at $\nu<E_0$ (see Figs.
\ref{fig:dos-interm}.a and \ref{fig:dos-interm}.c).
On the contrary, the  weighted scattering rate (lower panel of
Fig. \ref{fig:dos-interm}.a)
undergoes a drastic change  in the same temperature range. 
At $T=0.03\omega_0$, the latter consists of 
a pattern of equally spaced peaks at $\nu=E_0-n \omega_0$, 
whose intensity decays 
as $\nu$ moves away from the ground state energy. 
Retaining only the (largest) peak at $E_0-\omega_0$, we see that
the convolution integral ${\cal D}$ roughly follows the form of the 
spectral density $N^*(\omega+\omega_0-E_0)$, with a gap below
$\omega=\omega_0$ and a maximum around $\omega 
\approx  E_P$. 
The behavior changes at $T=0.2 \omega_0$, where the weighted
scattering rate acquires a gaussian distribution
centered at an energy $\sim E_P$ below the ground
state, so that
the maximum of ${\cal D}$ moves to $\omega\approx
2E_P$. 
Clearly, the transfer of spectral weight seen in 
 Fig.\ref{fig:sigma-interm}.a when going from $T=0.03\omega_0$ to $T=0.2\omega_0$ 
must be traced back to the 
change of behavior in the excitations at $\nu<E_0$ described here.

A more careful look at Fig. \ref{fig:dos-interm}.a (expanded in Fig.
\ref{fig:dos-interm}.b) 
shows that the change  in the overall distribution of $Im \Sigma 
e^{-\beta(\nu-E_0)}$ 
is accompanied by a slight shift of the position of the individual peaks. 
In fact,  as can be
shown by direct inspection of  
the continued fraction expansion of Ref.\cite{depolarone}, 
each  peak at $\nu<E_0$ is 
a replica of the  polaron band at $E_0$, of width $W$.
At temperatures $T\ll W$, due to the exponential weighting factor,  
only the low momentum states close to the bottom edges  contribute
(cf. Fig. \ref{fig:dos-interm}.b). Such states are
free-electron like, and give rise to an optical absorption spectrum 
with a weak coupling character.
As $T\gtrsim W$, on the other hand, 
more localized higher momentum states 
come into play,   and the polaronic behavior is recovered. 
We remark that this ``reentrant'' behaviour is not restricted 
to the zero density case treated here. 
It can be found also in a two site cluster \cite{sim0ne} as well as in the
Holstein model at half filling \cite{centroide}. The reason is ultimately due
to the different roles played by the  quantum and thermal fluctuations in the
polaron crossover region. Quantum fluctuations are known to stabilize the 
non-polaronic phase, while incoherent fluctuations such as static disorder or
thermal fluctuations stabilize the polaron. As the temperature increases, 
polarons are therefore first stabilized, 
before they eventually dissociate at a higher temperature.

\begin{figure}[htbp]
\centerline{\resizebox{7cm}{!}{\includegraphics{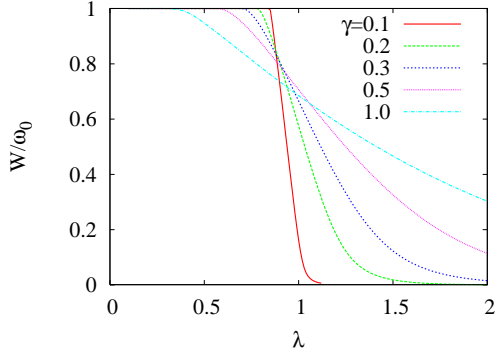}}}
\caption{(Color online) 
Width of the polaron band as a function of the interaction
  parameter $\lambda$ for different values of the adiabaticity
  parameter $\gamma$, at $T=0$.\label{fig:W}}
\end{figure}

According to the above arguments,
the transfer of spectral weight illustrated in Fig. \ref{fig:sigma-interm}.a 
is ultimately controlled by a temperature scale set by the
renormalized bandwidth $W$,
whose evolution
with the interaction strength is shown in
Fig. \ref{fig:W}.
In the strong coupling regime, where
$W$ vanishes exponentially,  
such phenomenon can hardly be observed in practice. 
On the other hand,
the very existence of a polaron subband separated from the continuum
of excited states 
requires moderately large values of $\lambda$, 
and the coexistence of electron-like states and localized states 
is specific to the adiabatic regime.
The competition between all these conditions 
explains why the low-temperature transfer of spectral weight described here 
occurs in  the vicinity of the adiabatic polaron crossover,
roughly for  $\lambda_c-0.5 \gamma \lesssim  \lambda \lesssim  \lambda_c+\gamma$
  and $\gamma\lesssim 0.5$ (shaded area in Fig.\ref{fig:PD}).

A formal expression for the optical conductivity in the limit $T\ll W$ is
derived in App. B, that we reproduce here:
\begin{eqnarray}
\label{sigma1-text}
\sigma(\omega) &=& [1-{\cal W}(0)] \frac{\pi}{\omega} \times \\ \nonumber
&\times  & 
\frac{\sum_{p=1}^{\overline{p}} 
        \int d\epsilon N(\epsilon)\phi(\epsilon)
                \rho^{>}(\epsilon,\omega+E_0-p\omega_0)
                        \overline{u}_p(\epsilon)}
{\sum_{p=1}^{\infty}\int d\epsilon N(\epsilon)
                        \overline{u}_p(\epsilon)}
\end{eqnarray}
where ${\cal W}(0)$ is a constant between $0$ and $1$ depending on the
interaction parameters, and the functions
$\overline{u}_p(\epsilon)$ are defined in Eq. (\ref{overlup}). This
formula replaces Eq. (7)  of Ref. \cite{BRoptcond}, which  was
incorrectly identified with the zero temperature limit of the optical 
absorption but is instead a preasymptotic contribution as $T\rightarrow 0$.

\subsection{Polaron Interband Transitions}

A detailed view of the low frequency optical absorption in the
intermediate coupling regime is presented in Fig. \ref{fig:sigma-interm}.b. 
As was mentioned in the previous Section,  the spectrum at the
lowest temperature  (full red curve) 
has a sharp threshold at $\omega=\omega_0$, followed by 
an absorption pattern which roughly reproduces the shape of the single
particle spectral density, shifted by $\omega_0$ 
[see the weak
coupling formula (\ref{eq:weak}) in the appendix, as well as  its
generalization Eq. (\ref{sigma1-text})]. 
For example, the peak at $\omega\simeq 0.17$ is directly related to
the maximum  observed in the spectral density at the top of the 
polaron band (compare Fig. \ref{fig:sigma-interm}.b 
with Fig. \ref{fig:dos-interm}.c), and its position
coincides with the expected value $\omega=\omega_0+W=0.173$. 
Upon increasing the temperature, 
the most striking effect in this low-frequency region is the 
emergence of a sharp asymmetric peak  \textit{within the gap}, whose
intensity follows a puzzling non-monotonic temperature dependence. 
Such resonance  has already been reported in 
Ref. \cite{BRoptcond}, where it was termed polaron interband
transition (PIT), and suggested as a possible explanation of the peaks
experimentally observed in Ref.\cite{TAIR}.
In order to understand its origin,
one has to go beyond the low-temperature 
arguments presented in the previous Section, and take into
account the contributions to the convolution integral in
Eq. (\ref{caldi}) coming from the states at $\nu>E_0$.

Because of thermal activation, there is a non-vanishing probability
for transitions involving an initial state at energies above the 
ground state. 
The Fermi golden rule tells us that
the transition probability at a given 
frequency $\omega$ will be proportional to the conjugate
density of the initial and final states separated by $\omega$, times
the occupation factor of the initial state.
Therefore, a sharp resonance  will arise if: i) 
many pairs of initial and final states exist that are separated
by approximately the same energy $\omega^*$, which is equivalent
to the condition that the energy dispersions in the initial and final subband 
are parallel; ii) such states are long lived, which is possible at
low temperatures because $Im \Sigma$
is exponentially suppressed in the whole region 
$E_0<\nu<E_0+\omega_0$. Both conditions can be fulfilled 
in the intermediate coupling regime,
as illustrated in  Fig. 3 of Ref.  \cite{BRoptcond}. 
Upon increasing the temperature,
the states involved in the transitions are eventually smeared out 
by thermal disorder,
which explains the non monotonic temperature dependence of this resonance.
\footnote{More generally, when more than one subband is present in the
  excitation spectrum,
  a fraction or even all of the corresponding states can have a 
vanishing scattering rate at low temperatures
(cf. Figs. 11 and 13 of Ref. \cite{depolarone}), and several PITs can
appear.}

This qualitative argument can be formalized as follows
(the detailed calculations are presented in  Appendix B).
When the scattering rate in the initial state is negligible, the
spectral function $\rho(\epsilon,\nu)$ can be replaced by a delta
function and the convolution integral  
in Eq. (\ref{caldi}) becomes [cf. Eq.  (\ref{DMM})]: 
\begin{eqnarray}
\label{diPITtxt}
{\cal D}(\omega) &\simeq & \int_{E_0}^\infty d\nu
N(\nu-Re\Sigma(\nu))\phi(\nu-Re\Sigma(\nu)) e^{-\beta(\nu-E_0)}  \nonumber\\
&\times& \left ( -\frac{1}{\pi}
Im \frac{1}{\omega+Re\Sigma(\nu)-Re\Sigma(\nu+\omega)+i\Gamma}\right ).    
\end{eqnarray}
This quantity is maximum 
 for those values of $\omega$ such that 
there is a wide range of energies $\nu$ where the denominator 
is close to zero.
This condition 
corresponds to the pair of equations
\begin{eqnarray}
 && \omega+Re\Sigma(\nu)-Re\Sigma(\nu+\omega) = 0 \nonumber\\
 && \left. \frac{\partial Re\Sigma}{\partial \nu} \right|_\nu
    -\left. \frac{\partial Re\Sigma}{\partial \nu} \label{resonancePIT}
    \right|_{\nu+\omega}
    =0
\end{eqnarray}
whose solution we denote as $\omega^*$ and $\nu^*$. The 
optical conductivity can be evaluated by expanding
Eq. (\ref{diPITtxt}) around these
special values, and taking $\Gamma\equiv -Im \Sigma(\nu+\omega)$ to be small
and constant in the frequency range of interest,  which gives 
\begin{eqnarray}
 & & \sigma_{PIT}(\omega) \simeq   
\frac{\pi}{{\sqrt{2Q\Gamma}\omega}}
\frac{(1-e^{-\beta\omega})e^{-\beta\nu^*}}{{\cal N}(\beta)} \times
 \\ 
& & \times N(\nu^*-Re\Sigma(\nu^*))
\phi(\nu^*-Re\Sigma(\nu^*))  
  {\cal F} [\frac{M(\omega-\omega^*)}{\Gamma}]. \nonumber
\label{PITsigappr}
\end{eqnarray}
where $M$ is the 
effective mass of 
the initial (and final) states and $Q$ is related to the curvature
of the dispersion. The function
\begin{equation}
  {\cal F}(y)=
\frac{\sqrt{y+\sqrt{y^2+1}}}{\sqrt{y^2+1}}
\end{equation}
 has an inverse square root tail  $\sim y^{-1/2}$ 
at large $y$, and tends to $1$  for $0<y\lesssim 1$, giving rise to a 
narrow peak of width $\Delta \omega \sim \Gamma/M$.  
Such asymmetric shape can be clearly identified in Fig.\ref{fig:sigma-interm}.b, at 
$\omega^*\simeq 0.3 \omega_0$.
Reminding that $\Gamma\sim e^{-\beta \omega_0}$, and using the
definition Eq. (\ref{sigma-compact}), 
we see that the weight of the narrow peak scales as
\begin{equation}
  w(\beta)\sim
  \frac{1-e^{-\beta\omega^*}}{\omega^*}\frac{e^{-\beta(\nu^*-E_0+\omega_0/2)}}
  {{\cal N}(\beta)}.
\end{equation}
This quantity
rises exponentially with temperature 
and then decreases as a power law at temperatures 
above some fraction 
of $\omega_0$, whose precise value depends on 
the parameters $\omega^*$, $\nu^*$ and $W$.

\section{Conclusion}

In this work, we have studied the optical absorption of electrons
interacting with phonons in the framework of
the Holstein model, 
applying a unified treatment
--- the dynamical mean-field theory --- which  
is able to account for both the quantum nature of the phonons and
the effects of a finite electronic bandwidth.
The present non-perturbative approach allows to span all the
ranges of parameters of the model, from weak to strong
electron-phonon interactions, 
and from the adiabatic to the antiadiabatic limit. 
The limits of validity of the standard available formulas are pointed out,
and the quantitative and qualitative deviations arising in wide 
ranges of the parameters space
are analyzed, which will hopefully lead to a better
understanding of the experiments performed on polaronic systems.

In the antiadiabatic regime, the present method allows to address the
evolution of the optical absorption  away from the 
single molecule limit, where the spectra reduce to a distribution of
delta functions. Considering a finite bandwidth $D$
yields an intrinsic width to the multi-phonon peaks,  which
is not uniform over the whole
spectral range (peaks at high frequency are broader than at low frequency) 
and generally does not obey the exponential narrowing predicted by 
the standard approaches. Moreover, in the strong coupling regime, 
the peaks are found to broaden 
 upon increasing the
temperature, in marked contradiction to the commonly accepted
results in the literature, which are based on uncontrolled approximations.

The situation is different in the adiabatic regime, 
where it is found that 
the nature of the optical absorption  changes drastically
at the sharp polaron crossover occurring at $\lambda\simeq \lambda_c$.
The weak coupling formula  (\ref{eq:weak-T0}) is
shown to extend qualitatively to finite values of $\lambda$, 
up to  the polaron crossover,
whose proximity is signaled by the emergence of a 
fine structure related to multi-phonon processes.
Beyond  the polaron crossover, 
 the absorption mechanism in the strong coupling regime is found to depend 
on an additional parameter,
the ratio $s/D$ between the broadening of the Franck-Condon line and
the free-electron bandwidth. The usual polaronic  gaussian lineshape described
by  Eq. (12) is recovered for $s\gg D$. In the opposite limit, 
$s\ll D$, a different absorption mechanism sets in, 
related to the photoionization of the polaron towards the free-electron
continuum. Correspondingly, the lineshape in this limit is entirely
determined by the shape of the free-electron band, and is
characterized by a sharp absorption edge at finite frequency,  
as described by Eq. (13).    
The optical spectra at intermediate values of $s/D$, a situation often
encountered in real systems, are not properly
described by any of the two limiting formulas Eqs. (12) and (13). 
In particular, 
due to finite bandwidth effects,
the frequency of the absorption maximum is appreciably lower than 
the usual estimate $\omega_{max}=2E_P$,  
and the lineshape is more asymmetric than the gaussian of
Eq. (12), which should be
taken into account when interpreting experimental data.

In the intermediate coupling region around $\lambda\simeq \lambda_c$, 
qualitatively new features arise that 
are distinctive of the adiabatic polaron crossover.
First, 
the optical absorption exhibits a reentrant behavior,  switching 
from weak-coupling like to polaronic-like upon increasing the temperature.
The temperature scale that governs such transfer of spectral weight is
of electronic origin, being
set by the renormalized polaronic bandwidth $W$, a quantity that 
is necessarily less  (or equal) than 
the phonon energy, and rapidly decreases upon increasing $\lambda$. 
In addition,  sharp peaks  with a
non-monotonic temperature dependence
emerge at typical phonon energies, and are most clearly visible in the region  
below the single-phonon absorption gap. These can be ascribed to
thermally excited resonant transitions between long lived states located in 
different subbands in the polaron
internal structure, and their peculiar temperature dependence 
results from the competition between the thermal activation and  
the thermal broadening of the corresponding states.

To conclude, let us observe that
the present formalism  can in principle be applied to any finite-dimensional
lattice by replacing $N(\epsilon)$ of Eq. (7) with the appropriate DOS.
It has already been  mentioned (cf. end of Section V A) that Eq. (13)
reproduces very accurately the one-dimensional exact diagonalization
data of Ref. \cite{ray} obtained for $s\ll D$. Even away from this limit, 
however, preliminary DMFT results obtained using 
the one-dimensional DOS $N(\epsilon)=1/ (\pi \sqrt{D^2-\epsilon^2})$ are in
qualitative agreement with the numerical results of
Ref. \cite{Fehske-opt}, at least in the high frequency part of the spectra.
In particular, in the intermediate coupling regime, 
the polaronic absorption maximum appears to be
shifted downwards as compared with the three-dimensional case 
(cf. Figs. 4 and 6 in Ref. \cite{Fehske-opt}). This result can be understood
from the  arguments presented in 
Section V A by observing that due to the singularity 
of the one-dimensional DOS, the function $N^*$ appearing in Fig. 
\ref{fig:dos-interm} is peaked around 
$\nu-E_0\simeq 0$ instead of $\nu-E_0\simeq E_P$. 

Concerning vertex corrections, which are absent in
our formalism, we have seen that these are irrelevant in the limiting
regimes of weak and strong coupling, as well as in the  
adiabatic regime $s \ll D$ at {\it any} coupling strength and 
for any lattice dimensionality (including in one
dimension).
On the other hand, their presence could quantitatively change the
optical spectra when 
both the phonon frequency and coupling strength lie in the intermediate regime.
From this point of view, the polaron interband transitions 
evidenced in the intermediate adiabatic crossover region, 
relying on the resonance condition Eq. (\ref{resonancePIT}),
could indeed be affected by the inclusion of vertex corrections.

As a last remark, we stress that the description of the optical properties  
provided in this work, based on the solution of the
Holstein model for a single particle, is valid in principle 
for a system of
independent polarons. 
How the above features are modified by
polaron-polaron interactions at finite densities remains a fundamental
question, that needs to be addressed in the future.

\section*{Appendix A: Derivation of limiting formulas}

\subsection*{Weak coupling limit, low temperatures $T\ll \omega_0$.} 
In the limit of weak interactions,
we can write the spectral function $\rho_\epsilon$ appearing 
in Eq. (\ref{caldi}) as 
$$
\rho_\epsilon = \rho^{(0)}_\epsilon+\delta \rho_\epsilon
$$
where  $\rho^{(0)}_\epsilon$ is non-interacting and reduces to a delta
function.
Substituting the previous expression in Eq. (\ref{caldi}) and taking into
account only the  contributions proportional to  $\delta\rho$ we get 
\begin{eqnarray}
\label{eq:Dpert}
  {\cal D}(\omega) &=&
 \int d\epsilon N(\epsilon) \phi(\epsilon)
 \int d\nu  e^{-\beta(\nu-E_0)}  \times  \\
          &  \times& \left [ \rho^{(0)}_\epsilon(\nu) \delta\rho_\epsilon(\nu+\omega) +
            \delta \rho_\epsilon(\nu) \rho^{(0)}_\epsilon(\nu+\omega)
           \right ]\nonumber
\end{eqnarray}
Performing the
two integrals in $\nu$ 
yields
\begin{eqnarray}
\label{eq:Dpert1}
  {\cal D}(\omega)  &=&
   -\frac{1}{\pi \omega^2}
 \int d\epsilon N(\epsilon) \phi(\epsilon)
  \left [
          e^{-\beta(\epsilon-E_0)} Im \Sigma(\epsilon+\omega)+\right.\nonumber\\
          & & \left. Im \Sigma(\epsilon -\omega) e^{-\beta(\epsilon-\omega-E_0)}
           \right ]
\end{eqnarray}
Perturbation theory gives, to  order $g^2$ and at zero density
\begin{equation}
\Sigma(\omega) = g^2 \left [ (1-n_B) G(\omega - \omega_0)+ n_B G(\omega + \omega_0)\right ]
\label{Sigma-pert}
\end{equation}
where $n_B$ is the Bose occupation number.
Substituting the self-energy into Eq. (\ref{eq:Dpert1})
gives rise to 4 terms, namely:
\begin{subequations}
\begin{eqnarray}
 {\cal D}(\omega)  &=& \frac{g^2}{\omega^2}\int d\epsilon  N(\epsilon)
 \phi(\epsilon) \times \nonumber \\
&\times  & \left [ (1-n_B) e^{-\beta(\epsilon-E_0)}  N(\epsilon+\omega-\omega_0)  \right. \label{eq:Dpert2_a1} \\
&+ &n_B e^{-\beta(\epsilon-E_0)}  N(\epsilon+\omega+\omega_0) \label{eq:Dpert2_a2} \\
&+ &(1-n_B) e^{-\beta(\epsilon-\omega-E_0)}  N(\epsilon-\omega-\omega_0) \label{eq:Dpert2_b1} \\
&+ &n_B e^{-\beta(\epsilon-\omega-E_0)}  N(\epsilon-\omega+\omega_0) \label{eq:Dpert2_b2} \left. \right ]
\end{eqnarray}
\label{eq:Dpert2}
\end{subequations}
At temperatures $T\ll\omega_0$,
the second and third term are
exponentially small, and can be neglected.
Note also that, since the above terms are explicitely proportional to $g^2$, 
the normalization factor (\ref{calenne}) can be evaluated in the
non-interacting limit, which gives 
$  \mathcal{N}(\beta) = (2/\beta) e^{-\beta D} I_1(\beta D)$,
where $I_1$ is the modified Bessel function of the first kind, whose 
limiting behaviors are  $\mathcal{N}\simeq  \sqrt{2/\pi}/(\beta
D)^{3/2}$ for $\beta D \gg 1$ and $\mathcal{N}\to 1$ for  $\beta D
\ll 1$.

At  temperatures much smaller
than the width of the band under study, 
due to the presence of the exponential factors, 
only the low energy edges of the density of states (corresponding to
low-momentum states) 
contribute to the integrals in Eq. (\ref{eq:Dpert2}). 
In this case we can make use of the following approximation:
\begin{equation}
\label{delta}
\int_\nu d\epsilon e^{-\beta (\epsilon-\nu) } (\epsilon -\nu)^\ell
f(\epsilon)   
\simeq 
\frac{f(\nu)}{\beta^{\ell+1}} \Gamma (\ell+1)
\end{equation}
where $\Gamma(n)$ is the Euler gamma function and
$f(\nu)$ is a generic continuous function.
By substituting the previous expression in equations
(\ref{eq:Dpert2}), only the term (\ref{eq:Dpert2}d) survives in the
limit $T\ll D$, and we obtain
\begin{equation}
\left. \sigma(\omega)\right|_{T\ll D}= \frac{g^2\pi}{\omega^3} 
 \phi(\omega-D-\omega_0)
N(\omega-D-\omega_0).
\label{eq:weak}
\end{equation}
which corresponds to  a sharp edge behavior  
$\sigma(\omega)\sim (\omega_0-\omega)^{3/2}$, in agreement with 
perturbative calculations 
in three dimensions\cite{Mahan}. Similar arguments lead to
$\sigma(\omega)\sim (\omega_0-\omega)^{d/2}$ if one assumes the appropriate
$d-$dimensional noninteracting band in Eq. (\ref{caldi}).

In the adiabatic regime, the condition $T\ll \omega_0$
ensures that $T\ll D$.
In the opposite antiadiabatic regime, 
the  result (\ref{eq:weak}) remains valid for
$T\ll D$, while for $D\ll  T \ll \omega_0$, we can set the exponential
factors equal to $1$ in Eqs. (\ref{eq:Dpert2}), leading to 
\begin{equation}
  \label{eq:weak-anti}
\left. \sigma(\omega)\right|_{D\ll T\ll \omega_0}
= \frac{g^2\pi}{\omega^3} F(\omega-\omega_0)
\end{equation}
where the function $F$  is defined as
\begin{equation}
  \label{eq:F}
  F(y)=\int_{-D}^{D-|y|} dx N(x)N(x+|y|)[\phi(x) +\phi(x+|y|)]
\end{equation}
(note that current vertex corrections are not included in this
formula; however, they lead to the same functional form in both the
adiabatic and anti-adiabatic limits).
In this temperature range, the  absorption related to the excitation 
of a single phonon  is nonzero in the range
$\omega_0-2D<\omega<\omega+2D$  and is symmetric around
$\omega=\omega_0$ (see the inset in Fig.  \ref{fig:anti}.a)
where $\sigma=(128/45\pi)g^2D/\omega_0^3$. 
By comparing Eq. (\ref{eq:F}) with Eq. (\ref{eq:weak}), we see that in the
antiadiabatic regime, the absorption peak
is shifted to lower frequency when the temperature is increased above
$T\simeq D$, and the sharp square root edge 
is converted into
$\sigma(\omega) \sim [\omega-(\omega_0-2D)]^3$.

\bigskip 
\subsection*{Strong coupling limit, narrow bands.} 
In the adiabatic limit ($\omega_0 \to 0$), the strong coupling regime
can be attained  by increasing $\lambda$ while keeping the polaron
energy $E_P$ finite.  
In this case, the spectral function of a single polaron becomes
independent of momentum (the $\epsilon$ index drops because the
bandwidth $D=E_P/\lambda\to 0$).  The self-energy in Eq. (2) becomes
$\Sigma=-g X$, where $X$ is a classical variable that acts as a source
of disorder that modifies the electronic levels according to the
Boltzmann distribution  \cite{depolarone,HoHu}.  
Correspondingly, the spectral function tends to a gaussian
\begin{equation}
  \rho(\nu)=\frac{1}{\sqrt{2 \pi s^2}}
  \exp\left[-\frac{\nu^2}{2s^2}\right]
\label{eq:strong}
\end{equation}
whose variance $s=\sqrt{2E_PT}$
is determined by the thermal fluctuations of the classical 
phonon field. 
Once the $\epsilon$-integral is factored out (yielding a numerical
prefactor),  the convolution
integral in Eq. (\ref{caldi}) is readily evaluated and yields
\begin{equation}
\label{eq:atomic-adiabatic}
  \sigma(\omega)_{s\gg D}=\frac{\pi D^2}{4\omega}\frac{1-e^{-\beta \omega}}
{\sqrt{4 \pi s^2}} 
\exp\left[-\frac{(\omega - 2E_P)^2}{4 s^2}\right]
\end{equation}
which coincides with the standard result of Refs. \cite{old,Reik,Mahan,Emin93}
It should be noted that, due to the initial assumption $\omega_0=0$,  the
above derivation is strictly valid only in the high temperature limit $T\gg
\omega_0$. However, 
Eq. (\ref{eq:atomic-adiabatic})  can be generalized
to all temperatures by setting
\begin{equation}
  s^2(T)=E_P \omega_0 \coth (\frac{\omega_0}{2T})
\end{equation}
[note that Eq. (\ref{eq:strong}) is also valid at
$T=0$, in which case the variance is determined by the phonon zero point
fluctuations and becomes $s=\sqrt{E_P \omega_0}$ \cite{depolarone,HoHu}].
Eq. (\ref{eq:atomic-adiabatic}) is valid in the strong coupling 
adiabatic regime provided that $s\gg D$.

Note that in the limit of an isolated molecule ($D=0$) at finite
$\omega_0$, which is qualitatively 
representative of the antiadiabatic regime,
the spectral function consists of a
distribution of delta functions, describing the multi-phonon
excitations inside the polaron potential-well\cite{Mahan}. 
The convolution integral in Eq. (\ref{caldi}) can be carried out, and
yields a discrete absorption spectrum, whose  \textit{envelope} 
in the strong coupling limit $\alpha^2=E_P/\omega_0\gg 1$ 
coincides with Eq. (\ref{eq:atomic-adiabatic}).

\subsection*{Strong coupling limit, wide bands.}

There is an alternative way of reaching the polaronic
adiabatic regime, i.e.  assuming that the phonon induced
broadening of electronic levels $s$ is negligible compared to the
electron dispersion $D$. For sufficiently large $E_P$, the optical 
absorption will be due to transitions from a perfectly
localized level whose electronic energy is $-2E_P$, to the
free-electron continuum. Put in the context of the convolution
integral  Eq. (\ref{caldi}), this corresponds to the following
replacements
\begin{eqnarray}
  \rho(\epsilon,\omega+\nu) &= & \delta(\epsilon-\omega-\nu)\\
  \rho(\epsilon,\nu)&=&\delta(\nu+2E_P) \label{eindep}
\end{eqnarray}
where the last quantity is independent of $\epsilon$.
 The result is
\begin{equation}
  \label{eq:Ray}
    \sigma(\omega)_{s\ll D\ll E_P}=\pi\frac{1-e^{-\beta \omega}}{\omega}
\phi(\omega-2E_P)N(\omega-2E_P)
\end{equation}
This result is 
equivalent to that  reported in Ref. \cite{ray,Firsov} for small
polarons in one-dimensional systems, and in Ref. \cite{Emin93} 
for large polarons in two and three dimensions.
However, the limit of a perfectly localized state assumed here 
is strictly valid only for $E_P\to \infty$. At finite $E_P$, a more accurate
formula can be obtained by 
restoring the $\epsilon$-dependence of the initial state
in Eq.(\ref{eindep}). 
This can be achieved by taking
\begin{equation}
  \rho(\epsilon,\nu)=-\frac{1}{\pi} Im \frac{1}{\nu - \Sigma(\nu) -\epsilon}
\simeq  -\frac{1}{\pi} \frac{Im  \Sigma(\nu)}{(\nu-\epsilon)^2} 
\end{equation}
where we have used the fact that both $Re \Sigma$ and $Im \Sigma$ are 
small around $\nu\simeq -2E_P$. The scattering rate at these energies
can in principle be extracted from the atomic limit of
Ref. \cite{Mahan}, because it is a
local quantity.  For $ s\to 0$ it is given by:
\begin{equation}
  -\frac{1}{\pi} Im \Sigma(\nu) e^{-\beta (\nu+E_P)} = \nu^2 \delta(\nu+2E_P) 
\end{equation}
Evaluation of Eq. (\ref{caldi})  now yields a more asymmetric shape
\begin{equation}
  \label{eq:Fratinov}
   \sigma(\omega)_{s\ll D}=\pi \frac{ 4E_P^2}{\omega^2}\frac{1-e^{-\beta\omega}}{\omega}
\phi(\omega-2E_P)N(\omega-2E_P)
\end{equation}
which reduces to the previous Eq. (\ref{eq:Ray}) in the limit
$E_P\gg D$.

It should be noted that when the appropriate one-dimensional DOS is
used, Eq. (\ref{eq:Fratinov}) describes much better the exact
diagonalization data of Ref. \cite{ray} than Eq. (\ref{eq:Ray})
itself, that was used by Alexandrov and coworkers in their Figs. 3.c
and 3.d.
A possible generalization of the present scheme to finite values of $s$
is currently under study.\cite{prepa}

\section*{Appendix B: Optical conductivity at low temperature}

In this appendix we give a formal expression for the optical
conductivity  in the limit of  low temperatures, which is useful for the
understanding of the results in the intermediate coupling regime.

We can formally separate  $\rho (\epsilon,\nu)$ in two parts, at energies
above and below the ground state
\begin{equation}
  \label{eq:rho-sep}
  \rho (\epsilon,\nu)=\theta(\nu-E_0)\rho^{>} (\epsilon,\nu)+
  \theta(E_0-\nu)\rho^{<} (\epsilon,\nu)
\end{equation}
[a similar separation will hold for
$N^*(\nu)$]
Accordingly, in  Eq. (\ref{sigma-compact}) we can separate in both 
numerator $\cal{D}$
and denominator $\cal{N}$ the contributions coming from energies lower
and higher than the ground state energy $E_0$,
obtaining
\begin{eqnarray}
{\cal D}(\omega)&=&{\cal D}^{>>}(\omega)+{\cal D}^{<>}(\omega)+{\cal
D}^{<<}(\omega) \label{D-sep}\\
{\cal N}&=&{\cal N}^{>}+{\cal N}^{<}
\label{N-sep}
\end{eqnarray}
In Eq. (\ref{D-sep}) the last contribution is negligible 
at low temperature for $\omega>0$ since 
$\rho^{<}(\omega+\nu)$ decays exponentially as $\exp[-\beta(E_0+\omega-\nu)]$.
On the contrary both ${\cal N}^{<}$ and ${\cal D}^{<>}(\omega)$ cannot be
neglected due to the thermal population factor $\exp [-\beta (\nu-E_0)]$ appearing
in Eq. 
(\ref{sigma-compact}). Let us calculate the remaining four terms separately.

\bigskip

\paragraph*{Calculation of ${\cal D}^{>>}(\omega)$}
\begin{eqnarray}
\label{def-DMM}
{\cal D}^{>>}(\omega)&=& \int_{E_0}^\infty d\nu \int_{-D}^D d\epsilon
N(\epsilon)\phi(\epsilon) e^{-\beta(\nu-E_0)} \times \nonumber\\
&\times& \rho^{>} (\epsilon,\nu) \rho^{>} (\epsilon,\omega+\nu)
\end{eqnarray} 
At low temperature, $\rho^{>} (\epsilon,\nu)$ weighted by the exponential factor
gets contributions only for $\nu\simeq E_0$.
 This
occurs when the temperature is so low that both the thermal
occupations of the first excited subband and of 
the incoherent background edge at $\nu=E_0+\omega_0$ are negligible,
i.e. for
$\beta \omega_0 \gg 1$. In this case the scattering
rate is negligible (the low-energy states are coherent), 
$\rho^{>} (\epsilon,\nu)$ can be replaced by 
$\delta [\nu-\epsilon-Re\Sigma(\nu)]$ and the integral in $\epsilon$ appearing
in Eq. (\ref{def-DMM}) can be carried out: 
\begin{eqnarray}
\label{DMM}
{\cal D}^{>>}(\omega) &=& \int_{E_0}^\infty d\nu 
N(\nu-Re\Sigma(\nu))\phi(\nu-Re\Sigma(\nu))  \nonumber\\
&\times&  e^{-\beta(\nu-E_0)}\rho^{>} (\nu-Re\Sigma(\nu),\omega+\nu).
\end{eqnarray}
At temperatures lower than the renormalized bandwidth $W$,
the thermal weighting factor selects an integration range 
of width $T$ around the ground state energy where, 
from Eqs. (\ref{def-semicirc}) and (\ref{def-vertex}), we have
\begin{eqnarray}
N(\nu-Re\Sigma(\nu))&\simeq& \frac{2}{\pi D^2} \sqrt{2D/Z} (\nu-E_0)^{1/2}
\label{NE0} \\
\phi(\nu-Re\Sigma(\nu))&\simeq& \frac{2 D /Z}{3 }  (\nu-E_0)
\label{phiE0}
\end{eqnarray} 
Here $Z=(1-\partial_\nu Re \Sigma)^{-1}$ is the ground state 
quasiparticle residue, equal to the inverse of the effective mass
in the case of a purely local self-energy.
Making use of Eq. (\ref{delta}) and the definition 
$E_0-Re\Sigma(E_0)=-D$, we obtain
\begin{equation}
  \label{eq-DMM}
  {\cal D}^{>>}(\omega) = C(T)
\left ( -\frac{1}{\pi} 
Im \frac{1}{\omega+E_0+D-\Sigma(\omega+E_0)}\right )
\end{equation}
with $C(T)=\sqrt{2/\pi} \ D^2/Z^{3/2} \ (T/D)^{5/2} $ for $T\ll W$.
In the strong coupling regime, the polaron band  becomes 
extremely narrow and 
can itself  be replaced by a delta function. For $T\gg W\to 0$, we obtain the
same expression Eq. (\ref{eq-DMM}) with 
$C(T)\simeq 4 W^2\to 0$, and  this
contribution becomes negligible.

\medskip

\paragraph*{Polaron interband transitions}

The previous formula for ${\cal D}^{>>}$
was derived assuming that $T\ll \omega_0$. 
Upon increasing the temperature, 
other contributions arise that involve states 
at frequencies $\nu$ away
from the ground state $E_0$. These can give rise to sharp resonances in the
optical absorption spectra (see Section IV B).  
Eq. (\ref{DMM}) can be rewritten as
\begin{eqnarray}
\label{diPIT}
&&{\cal D}^{>>}(\omega) = \int_{E_0}^\infty d\nu
N(\nu-Re\Sigma(\nu))\phi(\nu-Re\Sigma(\nu)) \\
&&\times  e^{-\beta(\nu-E_0)} \left ( -\frac{1}{\pi}
Im \frac{1}{\omega+Re\Sigma(\nu)-Re\Sigma(\nu+\omega)+i\Gamma}\right )
   \nonumber  
\end{eqnarray} 
where $\Gamma=-Im \Sigma(\nu+\omega)$ is taken to be small
and constant in the frequency range of interest.
A resonance will arise 
for those values of $\omega$ such that 
there is a wide range of energies $\nu$ where the denominator 
is close to zero. This condition corresponds to the
pair of equations
\begin{eqnarray*}
 && \omega+Re\Sigma(\nu)-Re\Sigma(\nu+\omega) = 0 \\
 && \left. \frac{\partial Re\Sigma}{\partial \nu} \right|_\nu
    -\left. \frac{\partial Re\Sigma}{\partial \nu}
    \right|_{\nu+\omega}
    =0
\end{eqnarray*}
whose solutions we denote $\omega^*$ and $\nu^*$. Now
the denominator in Eq. (\ref{diPIT}) can be expanded around these
special points, which gives 
\[
M(\omega-\omega^*)-Q(\nu-\nu^*)^2 +i \Gamma
\]
$M=1-(\partial\Sigma/\partial \nu)_{\nu^*}$ being the (equal) 
effective mass of
the initial (and final) states
and $Q$ a parameter related to the curvature
of $Re\Sigma$. If one assumes that the prefactors $N$ and $\phi$ are 
smooth functions in the
vicinity of $\nu^*$, these can be taken out of the
integral, which can be evaluated to 
\begin{eqnarray}
  {\cal D}^{PIT}&\simeq& \frac{N(\nu^*-Re\Sigma(\nu^*))
\phi(\nu^*-Re\Sigma(\nu^*))}{\sqrt{2 Q \Gamma}} \nonumber \\
&&\times  e^{-\beta\nu^*} 
  {\cal F} [M(\omega-\omega^*)/\Gamma]
\label{PITappr}
\end{eqnarray}
where the function
\begin{equation}
  {\cal F}(y)=
\frac{\sqrt{y+\sqrt{y^2+1}}}{\sqrt{y^2+1}}
\end{equation}
tends  to $1$ for $0<y\ll 1$ 
and has an inverse square root tail $\sim y^{-1/2}$ 
at large  $y$.

\medskip

\paragraph*{Calculation of ${\cal D}^{<>}(\omega)$} 
\begin{eqnarray}
\label{def-DmM}
{\cal D}^{<>}(\omega)&=& \int_{E_0-\omega}^{E_0} d\nu \int_{-D}^D d\epsilon
N(\epsilon)\phi(\epsilon) e^{-\beta(\nu-E_0)} \times \nonumber\\
&\times& \rho^{<} (\epsilon,\nu) \rho^{>} (\epsilon,\omega+\nu)
\end{eqnarray} 
The excitations in $\rho^{<}$ are mostly 
incoherent, being due to the presence of thermally activated phonons. This
is best expressed by showing explicitely the proportionality to the 
scattering rate
\begin{equation}
\label{rom}
\rho^{<}(\epsilon,\nu)= -\frac{1}{\pi} 
\frac{Im \Sigma(\nu)}{(\nu-Re \Sigma(\nu)-\epsilon)^2+(Im \Sigma(\nu))^2}
\end{equation}
Through Eq. (\ref{rom}), the product  
$e^{-\beta (\nu-E_0)}Im\Sigma(\nu)$ appears in Eq. (\ref{def-DmM}).
A sample plot of this product is reported in
Fig. \ref{fig:dos-interm}.b. 
At energies below the ground state, taking advantage of the 
continued fraction expansion of Ref.\cite{depolarone},
the imaginary part of $\Sigma$ can be  approximately written as a sum of
separate contributions  
\begin{equation}
\label{ImSm}
-Im\Sigma(\nu) = \sum_{p=1}^{\infty} e^{-\beta p \omega_0} u_p(\nu)
\end{equation}
As was the case for ${\cal D}^{>>}$, we treat separately the cases
$T\ll W$ and $T\gg W \to 0$. In the former case, 
$u_p$ is a function  with a 
square root edge at $\nu=E_0-p\omega_0$
\begin{equation}
\label{ImSmp}
u_p(\nu)= \overline{u}_p \ (\nu-E_0+p\omega_0)^{1/2}
\end{equation}
which is reminiscent of the behaviour of 
the spectral density at energy $E_0-(p-1)\omega_0$ and $\overline{u}_p$
is an unknown weighting factor.
As can be seen from Fig. \ref{fig:dos-interm}.b,  
each $u_p(\nu)$ has 
a bandwidth  which is independent on $p$, and equal to the width $W$ of the 
first polaronic band.  
Therefore,  when $T\ll W$, the square root edges can
all be replaced by delta functions by making use of Eq. (\ref{delta}).
When Eqs. (\ref{ImSm},\ref{ImSmp}) are substituted in Eq. (\ref{def-DmM})
the $\nu$-integral  can be carried out,  leading to
\begin{eqnarray}
\label{eq-DmM}
{\cal D}^{<>}(\omega)&=& \Gamma(3/2) T^{3/2} \sum_{p=1}^{\overline{p}} \int_{-D}^D d\epsilon
N(\epsilon)\phi(\epsilon)  
\times \nonumber\\
&\times& \rho^{>} (\epsilon,\omega+E_0-p\omega_0) \times \nonumber \\
&\times& \frac{1}{\pi}\frac{\overline{u}_p}{(E_0-p\omega_0-\epsilon -Re\Sigma(E_0-p\omega_0))^2} 
\end{eqnarray} 
where $\overline{p}=int(\omega/\omega_0)$.
In Eq. (\ref{eq-DmM}) we have neglected the exponentially vanishing 
scattering rate $Im\Sigma$ in the denominator of Eq.
(\ref{rom}).
At higher temperatures, or in the strong coupling regime, where
$\omega_0\gg T\gg W\to 0$, we can take
\begin{equation}
  u_p(\nu)=\bar{u}_p^\prime \delta(\nu-E_0-W+p\omega_0)
\end{equation}
which leads to an expression analogous to Eq. (\ref{eq-DmM}), with the
prefactor replaced by $e^{-\beta W}$ and $E_0\to E_0+W$, reflecting
the fact that the dominant weight is now carried by localized excitations
at the top of the band (see the discussion in Sec. III A). 

\medskip

\paragraph*{Calculation of ${\cal N}^{>}$.} 
\begin{equation}
\label{def-NM}
{\cal N}^{>}=\int_{E_0}^{\infty} d \nu \int_{-D}^{D} d\epsilon 
N(\epsilon) e^{-\beta (\nu-E_0)} \rho^{>} (\epsilon,\nu).
\end{equation}
Since a gap of width $\omega_0$ separates the coherent pole of 
$\rho^{>}$ from the incoherent continuum, 
for $\beta\omega_0 \gg 1$ only the lowest energy edge of the spectral density
contributes to the integral Eq. (\ref{def-NM}). Using
Eqs. (\ref{delta},\ref{NE0})  we get 
\begin{equation}
\label{eq-NM}
{\cal N}^{>}= \frac{2}{\pi D^2} \sqrt{2D/Z} \Gamma(3/2) T^{3/2}
\end{equation}

\medskip

\paragraph*{Calculation of ${\cal N}^{<}$.}
\begin{equation}
\label{def-Nm}
{\cal N}^{<}=\int_{-\infty}^{E_0} d \nu \int_{-D}^{D} d\epsilon 
N(\epsilon) \rho^{<} (\epsilon,\nu).
\end{equation}
As was stated before,  $\rho^{<}$ represents incoherent states as  given by  
Eq. (\ref{ImSm}). For $\beta W \gg 1$, the integrals in 
$\nu$ can be carried out 
with the help of Eq. (\ref{delta}) giving
\begin{eqnarray}
\label{eq-Nm}
{\cal N}^{<}&=& \Gamma(3/2) T^{3/2}
\sum_{p=1}^\infty \int_{-D}^{D} d\epsilon 
N(\epsilon) \times \nonumber \\
&\times& \frac{1}{\pi}
\frac{\overline{u}_p}{(E_0-p\omega_0-\epsilon-Re\Sigma(E_0-p\omega_0))^2}
\end{eqnarray}
At higher temperatures, $T\gg W$,  we get an analogous result, with the
prefactor replaced by $e^{-\beta W}$ and $E_0\to E_0+W$. 
By comparing the two expressions, it can be deduced that 
the term $ {\cal N}^{<}$  changes behavior
from $\sim T^{3/2}$ to $e^{-\beta W}$ at a temperature $T_W\sim
W/\log(D/W)$. The resulting normalization factor ${\cal N}$ has the form $\sim
T^{3/2}$ in the weak coupling regime and $\sim e^{-\beta W}\to 1$ in the
strong coupling regime. In the intermediate coupling regime, where
$W$ is of the order of $\omega_0$, a crossover between the two
behaviors occurs at $T_W$. 

\bigskip

\paragraph*{Final formula for $\sigma(\omega)$ at $T=0$.} 
Collecting the results Eqs. 
(\ref{eq-DMM},\ref{eq-DmM},\ref{eq-NM},\ref{eq-Nm}) into
Eq. (\ref{sigma-compact}) we get the final result.  
To separate the terms (\ref{eq-DMM}) and (\ref{eq-NM}) 
which involve the coherent states close to the ground state $E_0$
from the remainder, it is useful to introduce a temperature dependent weight 
\begin{equation}
{\cal W}(T) = \frac{{\cal N}^>}{{\cal N}}
\end{equation}
which tends to a constant for $T\ll W$, and to a (different) constant 
for $T\gg T_W$.
Then
\begin{equation}
\sigma (\omega) = {\cal W}(T) \sigma_0(\omega) + \left[1-{\cal W}(T)\right] \sigma_1(\omega)
\end{equation}
where, from Eqs. (\ref{eq-DMM},\ref{eq-NM}):
\begin{equation}
\label{sigma0}
\sigma_0(\omega) = \frac{\pi D/Z}{\omega}    T \left ( -\frac{1}{\pi}Im
\frac{1}{\omega+E_0+D-\Sigma(E_0+\omega)}\right )
\end{equation}
Eq. (\ref{sigma0}) is Eq. (7) of Ref. \cite{BRoptcond}, where it was
incorrectly identified with the zero temperature limit of the optical 
absorption (note that the linear dependence on 
temperature makes this contribution vanish at $T=0$).

Under the above hypothesis the remaining term $\sigma_1$ does not depend
explicitely on temperature. In fact, using Eqs. (\ref{eq-DmM},\ref{eq-Nm}) 
we get a formal expression
\begin{equation}
\label{sigma1}
\sigma_1(\omega) = \frac{\pi}{\omega} 
\frac{\sum_{p=1}^{\overline{p}} 
        \int d\epsilon N(\epsilon)\phi(\epsilon)
                \rho^{>}(\epsilon,\omega+E_0-p\omega_0)
                        \overline{u}_p(\epsilon)}
{\sum_{p=1}^{\infty}\int d\epsilon N(\epsilon)
                        \overline{u}_p(\epsilon)}
\end{equation}
where 
\begin{equation}
\label{overlup}
\overline{u}_p(\epsilon) = 
\frac{\overline{u}_p}{(E_0-p\omega_0-\epsilon -Re\Sigma(E_0-p\omega_0))^2}
\end{equation}
and at $T=0$ we have $\sigma(\omega)=[1-{\cal W}(0)]\sigma_1(\omega)$.
Note that, since the sum is limited to $\bar{p}$,  at low temperature
only the first term in the  numerator of Eq. (\ref{sigma1})
contributes to the Drude peak at $\omega\simeq 0$ and to the 
one-phonon threshold at $\omega=\omega_0$ .

To recover the perturbative expression Eq. (\ref{eq:weak}) we consider 
only a single contribution in the sums appearing
in Eq. (\ref{sigma1}). 
Then we perform the integral in
the numerator of Eq. (\ref{sigma1}) using the 
free spectral function 
$\rho^{>}(\epsilon,\omega+E_0-p\omega_0)=\delta(\omega+E_0-p\omega_0-\epsilon)$
obtaining
\begin{equation}
\label{sigma1p}
\sigma_1(\omega) = \frac{\pi}{\omega} 
\frac{
        N(\omega+E_0-\omega_0)\phi(\omega+E_0-\omega_0)
                        \overline{u}_1(\omega+E_0-\omega_0)}
{\int d\epsilon N(\epsilon)
                        \overline{u}_1(\epsilon)}.
\end{equation}
To proceed further we have from Eq. (\ref{Sigma-pert})
\begin{equation}
\overline{u}_1(\omega+E_0-\omega_0)\simeq -2 g^2 \frac{2\sqrt{2D}}{D^2\omega^2}
\end{equation}
where we have neglected the self-energy term in the denominator of 
$\overline{u}_1$ for $\beta \omega_0\gg 1$.
Eq. (\ref{eq:weak}) follows by multiplying (\ref{sigma1p}) by the factor 
$(1-{\cal W})$.

\end{document}